\documentclass[ amsmath,amssymb,aps, reprint]{revtex4-2}

\usepackage{graphicx}

\usepackage{bm}
\usepackage[colorlinks=true,citecolor=red,linkcolor=blue,urlcolor=blue]{hyperref}

\begin{document}

\title{Odd slip at chiral active surfaces}

\author{Yuto Hosaka}
\email{hosaka.yuto.7r@kyoto-u.ac.jp}
\affiliation{Max Planck Institute for Dynamics and Self-Organization (MPI-DS), Am Fassberg 17, 37077 G\"{o}ttingen, Germany}
\affiliation{Department of Mathematics, Kyoto University, 606-8502 Kyoto, Japan}

\author{Andrej Vilfan} 
\email{andrej.vilfan@ijs.si}
\affiliation{Jo\v{z}ef Stefan Institute, Jamova 39, 1000 Ljubljana, Slovenia}

\date{\today}

\begin{abstract}
Odd slip at solid-liquid interfaces, characterized by a velocity component perpendicular to the applied shear stress, introduces a novel class of odd transport phenomena with transverse momentum transfer at boundaries, distinct from odd viscosity, odd elasticity, or odd diffusivity. This effect arises on chiral active surfaces with persistent energy input and defined handedness, such as carpets of rotating biological or synthetic cilia with flexible anchoring.  Although odd slip itself is dissipationless, its maintenance in active systems comes at an energetic cost, which we estimate from the microscopic model.  With respect to its orientation dependence, odd slip can be classified as pseudo-scalar or pseudo-vectorial. We use a perturbative expansion to determine the resistance tensors of problems with different levels of broken symmetries. They reveal how the surface chirality and body asymmetry together violate the Onsager symmetry, but uphold the Onsager-Casimir reciprocity. Strikingly, suspensions of particles with odd slip embedded in a conventional Newtonian bulk fluid exhibit odd viscosity, providing a new minimal mechanism for its emergence beyond current microscopic models.
\end{abstract}

\maketitle

\section{Introduction}

Chirality --- the distinction between left- and right-handed configurations --- pervades nature from molecular scales to the bodies of organisms and the structure of their colonies \cite{wagniere2007chirality,Inaki.Matsuno2016,ArandaDiaz.Salama2021}. Examples span from enantiomeric molecular species that determine chemical reactivity and sensory perception to the helical architectures of biomolecules and the rotational motion of microorganisms~\cite{lauga2006swimming, lettermann2025chirality}.

\begin{figure*}
  \centering
  \includegraphics[width=17cm]{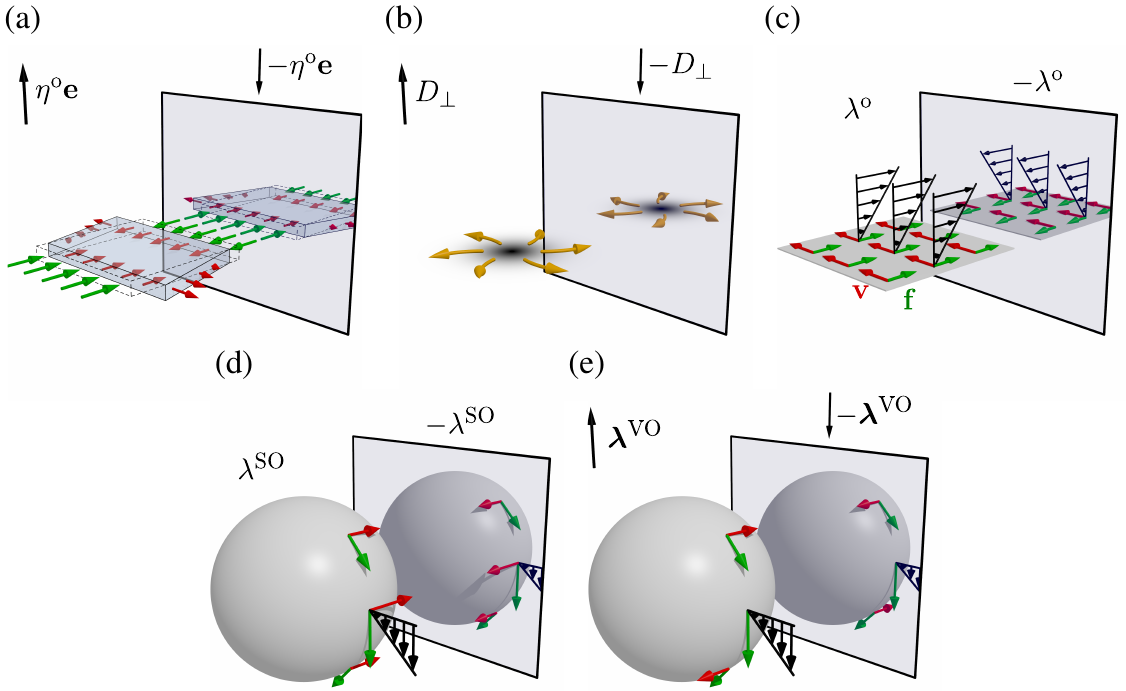}
  \caption{\label{fig:1}Parity-violating phenomena in a fluid. (a) Odd viscosity: a stress field (green) induces a skewed shear flow (red). (b) Odd diffusivity: the mean particle flow has a direction that is rotated with respect to the density gradient. (c) Odd slip: the slip velocity on the surface (red) is perpendicular to the applied shear stress (green). In each of the three cases, the response changes sign in the mirror image, corresponding to a parity transformation. (d) Pseudo-scalar odd slip is independent of the surface orientation. (e) Pseudo-vector odd slip changes direction depending on the product $\boldsymbol\lambda^\text{VO}\cdot \mathbf{\hat n}$.}
\end{figure*}

A recent approach to modeling chiral systems within continuum theories introduces anti-symmetric components into material modulus tensors that govern transport phenomena across various media, including fluids, elastic solids, diffusive systems \cite{fruchart2023odd} and even human crowds \cite{Gu.Bartolo2025}. These anti-symmetric components give rise to striking effects: dissipationless odd viscosity (Fig.~\ref{fig:1}a) generates transverse hydrodynamic responses \cite{avron1998,banerjee2017,soni2019odd} or anomalies from the ordinary Hall effect in electron fluids \cite{berdyugin2019measuring}; odd elasticity enables adaptive locomotion in active solids \cite{scheibner2020odd,tan2022odd,Veenstra.Coulais2025}; and odd diffusivity (or its inverse, mobility) leads to enhanced self-diffusion \cite{hargus2021odd,kalz2022collisions,Kalz.Sharma2024} (Fig.~\ref{fig:1}b). 
In terms of dimensionality, odd viscosity and elasticity manifest in two- or three-dimensional continua, whereas odd diffusivity is a property of point particles (systems of zero dimension or scalar field) in an external medium.
This raises the question of whether odd phenomena can also emerge in other spatial dimensions.

Here, we identify a previously unexplored route to parity breaking at active fluid-solid interfaces: an odd slip boundary condition defined through a chiral surface velocity (Fig.~\ref{fig:1}c). Depending on whether the chirality is intrinsic to the surface or induced by an external field, its nature can be pseudo-scalar (Fig.~\ref{fig:1}d) or pseudo-vectorial (Fig.~\ref{fig:1}e). We propose a microscopic realization of odd slip arising on ciliated surfaces, where flexible cilia get deflected by the external flow, while their intrinsic sense of active rotation breaks the chiral symmetry.  Finally, we show that in metafluids containing chiral-slip particles, odd viscosity emerges naturally even in the dilute limit, without the need for considering two-body effects. These findings establish chiral boundary slip as a fundamental mechanism for realizing odd hydrodynamics and provide new design principles for active and parity-broken materials.

\section{Odd slip}

We consider an interface between a solid object and a Newtonian fluid at zero Reynolds number. The fluid phase is described by the Stokes equation and the incompressibility condition
\begin{equation}
  \label{eq:stokes}
  \nabla \cdot \bm{\sigma}=\bm{0},\qquad   \nabla\cdot\mathbf{v}=0\,,
\end{equation}
where the stress tensor is given by $\boldsymbol{\sigma}=-p\mathbf{I}+\eta(\nabla\mathbf{v}+\nabla\mathbf{v}^\top)$ with the pressure field $p$, the velocity field $\mathbf{v}$, and the shear viscosity $\eta$.

The vast majority of microhydrodynamic problems assume the no-slip boundary condition at the fluid-solid interface, i.e., $\mathbf{v}=\bm{0}$ in the co-moving frame where the velocity of the solid surface is zero. However, flows at nanoscales or along superhydrophobic surfaces can allow a non-zero tangential slip velocity, which is proportional to the shear stress at the boundary \cite{LAUGA.STONE2003}. The classical Navier slip, also called partial slip, condition states $\mathbf{v} =(\lambda/\eta)  \left(\mathbf{I}-\mathbf{\hat n \hat n}\right)\cdot \boldsymbol{\sigma}\cdot\mathbf{\hat n}$, where $\lambda$ represents the slip length and $\mathbf{\hat{n}}$ is a surface normal pointing into the fluid.

On a chiral active surface, the linear response can take the form
\begin{align}
	\mathbf{v} =
         \frac{\lambda^\text{e}}{\eta}
        \left(\mathbf{I}-\mathbf{\hat n \hat n}\right)
        \cdot(\boldsymbol{\sigma}\cdot\mathbf{\hat n})
	   +  
       \frac{\lambda^\text{o}}{\eta}
       \mathbf{\hat n}\times
	   (\boldsymbol{\sigma}\cdot\mathbf{\hat n}),
    \label{eq:chiralBC}
\end{align}
where $\lambda^\text{e}$ is the classical or even and $\lambda^\text{o}$ 
the chiral or odd slip length.  More generally, Eq.~\eqref{eq:chiralBC} can be rewritten as $v_i=\mathcal{M}_{ij}\sigma_{jk}{\hat n}_k$, which also allows a direction-dependent slip \cite{BAZANT.VINOGRADOVA2008,Korneev.Abanov2021}. Because the slip velocity can only have a tangential direction and normal stress does not induce any slip, $\boldsymbol{\mathcal{M}}$ has to satisfy the conditions $\mathcal{M}_{ij}{\hat n}_j=0$ and ${\hat n}_i \mathcal{M}_{ij}=0$. The matrix $\bm{\mathcal M}$ can be decomposed into a symmetric and an anti-symmetric part $\mathcal M_{ij}=\mathcal M ^\text{e}_{ij} + \mathcal M ^\text{o}_{ij}$, with $\mathcal M ^\text{e}_{ij}=\mathcal M ^\text{e}_{ji}$ and $\mathcal M ^\text{o}_{ij}=-\mathcal M ^\text{o}_{ji}$. 
The slip defined in Eq.~\eqref{eq:chiralBC} is described by the matrices $\mathcal{M}_{ij}^\text{e}=(\lambda^\text{e}/\eta)(\delta_{ij}-\hat n_i \hat n_j)$ and $\mathcal{M}_{ij}^\text{o}=(\lambda^\text{o}/\eta) \epsilon_{ikj} \hat n_k$.
The dissipation density in the boundary is $\hat{\mathbf{n}}\cdot \boldsymbol{\sigma} \cdot \mathbf{v}=\hat{\mathbf{n}}\cdot \boldsymbol{\sigma} \cdot \boldsymbol{\mathcal M} \cdot \boldsymbol{\sigma} \cdot \hat{\mathbf{n}}$ and only contains the even part $\boldsymbol{\mathcal M}^\text{e}$. Like odd viscosity, the odd slip is dissipationless, even though it can influence the dissipation indirectly through its effect on the flow pattern.

The odd slip can, especially if driven by an external field, in principle have any dependence on the direction of the surface normal $\mathbf{\hat n}$. We represent this dependence with a multipole expansion,
\begin{equation}
  \label{eq:multipole}
\lambda^\text{o}(\mathbf{\hat n}) = \lambda^\text{SO} + \boldsymbol{\lambda}^\text{VO} \cdot\mathbf{\hat n}+ \ldots
\end{equation}
which we cut off after the dipole order. Because the odd slip length changes sign under inversion, $ \lambda^\text{SO}$ has the transformation properties of a pseudo-scalar (SO: scalar odd) and $ \boldsymbol \lambda^\text{VO} $ the properties of a pseudo-vector (VO: vectorial odd). Typically, the pseudo-scalar slip describes response that is intrinsic to the surface (Fig.~\ref{fig:1}d), whereas pseudo-vectorial slip is externally driven or controlled (Fig.~\ref{fig:1}e), for example by a magnetic field that is rotating around the direction specified by $ \boldsymbol \lambda^\text{VO} $.

\section{Flexible cilia model leads to effective odd slip}

One possibility to realize odd slip on a surface is to cover it with actively rotating cilia. In biology, 
motile cilia are hair-like cellular organelles that transport the surrounding fluid by moving periodically along a non-reciprocal path \cite{gilpin2020multiscale}. Ciliary beating is powered by the hydrolysis of ATP, therefore subject to persistent energy input that breaks the time-reversal symmetry. Ciliary beating patterns in general show at least some degree of chirality \cite{Striegler.Geyer2025} whereby their sense of rotation depends on the species and the type of cilia. Therefore, they also provide the broken parity symmetry needed for the odd slip. 
Most cilia show a strong asymmetry between working and recovery stroke, which allows them to move the fluid efficiently. However, some cilia can also move along conical trajectories with a small tilt, e.g., those in the left-right organizer \cite{Ferreira.Vermot2017}. 
Likewise, artificial biomimetic cilia are frequently actuated by a rotating magnetic field, which in the simplest case follows the mantle of a tilted cone \cite{Vilfan.Babic2010}. 
We will use conical trajectories without a tilt in our model in order to avoid any additional active flows that appear in the absence of external stress.
Although ciliary flows are complex and chaotic \cite{Supatto.Vermot2008}, they can be coarse-grained to the description with an effective slip velocity if the ciliary layer is thin in comparison to the dimension of the object \cite{Osterman.Vilfan2011}.

\begin{figure*}
  \centering
  \includegraphics[width=17cm]{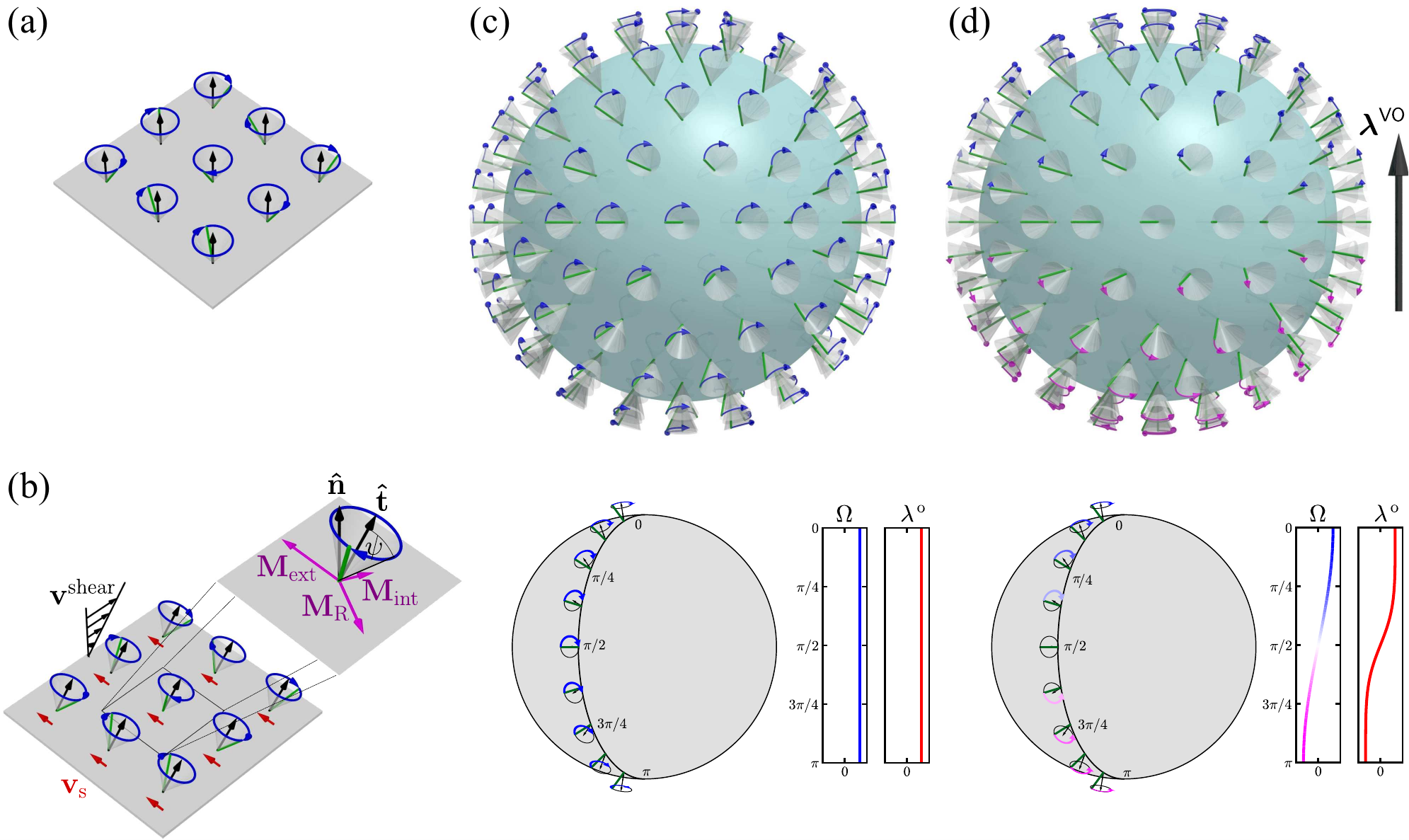}
  \caption{\label{fig:2}
    Flexible cilia cause an effective odd slip velocity. (a) Without external flow, cilia rotate symmetrically around the surface normal and do not produce any long range flow. (b) In the presence of an external shear, cilia get deflected and produce an effective slip flow (red arrows; only the odd component is shown). The inset shows the in-plane components of the three torques acting on the cilium (magenta). (c) Pseudo-scalar odd slip emerges when all cilia on a surface rotate in the same sense. (d) Pseudo-vectorial odd slip emerges when the direction of rotation is given by an external vector.
    The schematics in panels (c) and (d) illustrate the rotation of cilia along a meridional line on a sphere with clockwise (anti-clockwise) rotation represented by blue (magenta) arrows. The plots show the variation of the angular frequency $\Omega$ and odd slip length $\lambda^\text{o}$ along the meridian.
  }
\end{figure*}

We model a flexible cilium as follows. The cilium is described as a thin rod of length $L$, rotating with angular frequency $\Omega$ around an axis given by a unit vector $\mathbf{\hat t}$ (Fig.~\ref{fig:2}a). A deflection of the axis $\mathbf{\hat t}$ from the surface normal ($\mathbf{\hat n}$) is countered by a restoring torque
\begin{equation}
  \label{eq:restoring}
  \mathbf{M}_\text{R}=K \mathbf{\hat t} \times \mathbf{\hat n}
\end{equation}
with an angular elastic constant $K$. In addition, the cilium is subject to torques caused by the hydrodynamic drag due to its own motion and due to the external shear flow. We assume that the deflections of a cilium are sufficiently overdamped that the hydrodynamic torques can be averaged over one beating cycle.

With an externally imposed shear stress $\mathbf{f}=\bm\sigma \cdot\mathbf{ \hat n}$ on the surface, the velocity profile above the surface is $\mathbf{v}^\text{shear}=\mathbf{f} z/\eta$. For a small deflection ($\mathbf{\hat t}\approx\mathbf{\hat{n}}$), the mean torque acting on a cilium exposed to this shear flow is
\begin{equation}
  \label{eq:mext}
   \mathbf{M}_\text{ext}= \frac {L^3 C_N \cos^2\psi}{3\eta} \mathbf{\hat n} \times \mathbf{f}\;,
\end{equation}
with $\psi$ being the semi-cone angle measured from the direction $\mathbf{\hat{t}}$. We have modeled the hydrodynamics using the resistive force theory, which assumes that the flow is not perturbed by the presence of the cilium and that the force exerted per unit length is proportional to the local relative fluid velocity, multiplied by the drag coefficients $C_N$.

In addition, the rotation of the cilium around the axis of the cone leads to a hydrodynamic drag
\begin{equation}
  \label{eq:Mint}
   \mathbf{M}_\text{int}= \frac 13 L^3 C_N \sin^2\psi \, \Omega \mathbf{\hat t}\;.
\end{equation}

Whereas the normal component of the torque is counterbalanced by the active propulsion mechanism, the tangential components need to be counterbalanced by the restoring torque (Fig.~\ref{fig:2}b), which leads to the torque balance condition
\begin{equation}
  \label{eq:tbalance}
  \mathbf{\hat n} \times ( \mathbf{M}_\text{R}+ \mathbf{M}_\text{ext}+ \mathbf{M}_\text{int})=\mathbf{0}\;.
\end{equation}
The above equation is solved by
\begin{equation}
  \label{eq:tsolution}
\mathbf{\hat n}\times  \mathbf{\hat t}=\frac{\cot^2\psi}{\Omega \eta} \frac{\tilde K \mathbf{\hat n}\times \mathbf{f}+ (\mathbf{I}-\mathbf{\hat n}\mathbf{\hat n})\cdot \mathbf{f}}{\tilde K^2 + 1}
\end{equation}
where we introduced the dimensionless stiffness as
\begin{equation}
  \label{eq:K}
  \tilde K=\frac{3 K}{\Omega L^3 C_N \sin^2 \psi}\,.
\end{equation}

A tilted cilium changes its distance from the plane between the ``working'' and the ``recovery stroke'', which results in a net flow. The flow can be characterized by the volume flow rate $Q$, i.e., the fluid volume passing through a vertical half-plane per unit time. The flow rate can also be expressed as a vector $\mathbf{Q}$ with the direction corresponding to the direction of the fluid transport and can be calculated as follows.
A tangential point force $\mathbf F$ at distance $z$ from the surface induces a long-range flow with the flow rate $\mathbf{Q}=\frac 1 {\pi \eta} z \mathbf{F}$ \cite{Smith.Gaffney2008}. 
We use the force density from the resistive force theory, $C_N \dot {\mathbf{x}}$, where $\dot{\mathbf{x}}$ is the instantaneous velocity of a point on the cilium that follows the mantle of a tilted cone.
After integrating it over the length of the cilium and averaging it over one period of motion, we obtain the mean volume flow rate \cite{Smith.Gaffney2008}
\begin{equation}
  \label{eq:Q}
  \mathbf{Q} = \frac{L^3 C_N \Omega}{6\pi\eta} \sin^2\psi \mathbf{\hat n} \times \mathbf{\hat t}\,.
\end{equation}

A carpet of cilia with an area density $\rho$ and a uniform tilt produces a fluid flow with a velocity that becomes homogeneous at a distance of a few ciliary lengths above the surface. This velocity can be expressed as 
$\mathbf{v}_\text{S}=\pi \rho \mathbf{Q}$ 
and can be treated as an effective slip velocity on the surface \cite{Osterman.Vilfan2011}.
It contains a component parallel to $\mathbf{f}$, corresponding to an even or classical Navier slip, which we will not follow further.
The odd slip is given by the velocity component that is normal to $\mathbf{f}$ and reads
\begin{equation}
  \label{eq:vsn}
  \mathbf{v}_\text{S}^\perp=\frac{\rho L^3 C_N \cos^2\psi}{6 \eta^2} \frac{\tilde K}{\tilde K^2+1} \mathbf{\hat n} \times \mathbf{f}\;.
\end{equation}
Therefore, the surface covered in flexible cilia can be treated as an effective odd Navier slip with the odd slip length
\begin{equation}
  \label{eq:lambdaodd}
  \lambda^\text{o}=\frac{\rho L^3 C_N \cos^2\psi}{6 \eta} \frac{\tilde K}{\tilde K^2+1}\;.
\end{equation}
The equations show that the dependence on $\tilde K$ is non-monotonic. The odd slip length naturally vanishes for very stiff cilia (when no deflection occurs), but also for too flexible cilia, where the deflection becomes perpendicular to the shear flow. The maximum odd slip is obtained with $\tilde K=1$. 
With parameter values of $\psi=30^\circ$, $C_N=1.2\pi \eta$ and $\rho=L^{-2}$, the attainable odd slip length is of the order $\lambda^\text{o}\approx 0.25 L$. 

For internally driven cilia, the angular frequency $\Omega$ is independent of the orientation of the surface normal, and the resulting odd slip is pseudo-scalar in its nature (Fig.~\ref{fig:2}c). If the cilia are rotated by an external field, the dependence is $\Omega =\mathbf{\Omega}^E  \cdot \mathbf{\hat n}$, and likewise $\lambda^\text{o}=\bm{\lambda^\text{VO}}\cdot\mathbf{\hat n}$, leading to a pseudo-vectorial odd slip (Fig.~\ref{fig:2}d). 

Artificial magnetic cilia built on elastic filaments and driven by a rotating magnetic field, such as those reported in Ref.~\cite{Shields.Superfine2010}, are a candidate for pseudo-vectorial odd slip. With a length of $L=25\,\rm \mu m$, density of $0.02\,\rm \mu m ^{-2}$, and $\psi=6^\circ$, one can achieve odd slip length of up to $\lambda^{\rm o}\sim L$.

Finally, we note that in this section we assumed that the ciliary carpet is sparse enough to neglect the hydrodynamic coupling between cilia, which are only affected by the imposed external flow. However, most ciliated surfaces are dense enough that hydrodynamic interactions lead to the synchronization of cilia and the emergence of metachronal waves \cite{vilfan2006hydrodynamic}. This coupling introduces spatio-temporal variation in the response to the flow and, even in the time average, the response becomes intrinsically non-linear. 
We expect the linear response to weak perturbations to follow the same mechanism qualitatively, although the resulting slip length will be affected quantitatively.

\subsection{Energetic cost of maintaining an odd slip}

As shown above, the odd slip itself is formally dissipationless because it is orthogonal to the applied stress. Here, dissipationless means that it does not require a flux of work from the fluid into the surface layer, as ordinary Navier slip does. Nevertheless, the active driving mechanism of cilia demands a constant energy input. It is therefore instructive to compare the effect of slip to the dissipated power. The dissipation density is determined by the torque (Eq.~\eqref{eq:Mint}), multiplied by the angular velocity and the area density of cilia:
\begin{equation}
    P_\text{diss}=\frac{\rho L^3 C_N \Omega^2 \sin^2\psi }{3}\,.
\end{equation}
We estimate that the ciliary carpet is functional until the inclination reaches $30^\circ$, which corresponds to $\left|\hat{\mathbf{n}}\times \hat{\mathbf{t}}\right|=0.5$. At stronger flows, the nonlinearities dominate and eventually the ability to produce odd slip breaks down. By combining Eqs.~\eqref{eq:tsolution} and \eqref{eq:vsn} while assuming $\tilde K=1$, we obtain the estimate 
\begin{equation}
    (v_\text{S}^\perp f)_\text{max}= (v^\perp_\text{S-max})^2 \eta /\lambda^\text{o}=\frac{\tan^2\psi}{8} P_\text{diss}\,.
\end{equation}
In other words, the constant dissipation density exceeds the product of shear stress and odd slip by about a factor of 24. Although the estimated dissipation is specific to the ciliary model, our calculations illustrate how continuous energy input associated with ciliary activity can sustain odd slip in this particular realization.
It takes a significant amount of dissipation to produce ``dissipationless'' slip.

\section{Rigid body resistance}

The broken symmetries of the resistance matrix arise from an interplay between broken symmetries in the body shape, the fluid, and, in our case, also in the interface between them \cite{khain2024trading,hosaka2024chirotactic,everts2024dissipative}.  For a rigid object of any shape, the linear relation 
$   \bm{\mathsf{F}}= - \bm{\mathsf{R}}\cdot\bm{\mathsf{V}}$
connects translational and angular velocities and the strain-rate tensor (in short $\bm{\mathsf{V}}= [\mathbf{V}\, ,\boldsymbol{\Omega}\, ,{-\mathbf{E}^\infty}]$) and the hydrodynamic force moments (in short $\bm{\mathsf{F}}= [\mathbf{F}\,,\mathbf{T}\, , \mathbf{S}]$) via the grand resistance tensor
\begin{align}
    \bm{\mathsf{R}}
    =
    \begin{pmatrix}
        \mathbf{R}^{FV} & \mathbf{R}^{F\Omega} & \mathbf{R}^{FE} \\
        \mathbf{R}^{TV} & \mathbf{R}^{T\Omega} & \mathbf{R}^{TE} \\
        \mathbf{R}^{SV} & \mathbf{R}^{S\Omega} & \mathbf{R}^{SE} 
    \end{pmatrix}
    .
    \label{eq:grand}
\end{align}
In this section we will discuss the symmetry properties of the grand resistance tensor of a body with odd slip, derive a general perturbative solution, and compute it for three minimal problems with different origins of broken symmetry.
We start by deriving the Lorentz reciprocal theorem with slip boundary conditions and then use it to derive the symmetry properties and the perturbative solution.

\begin{figure*}
\centering
\includegraphics[width=16cm]{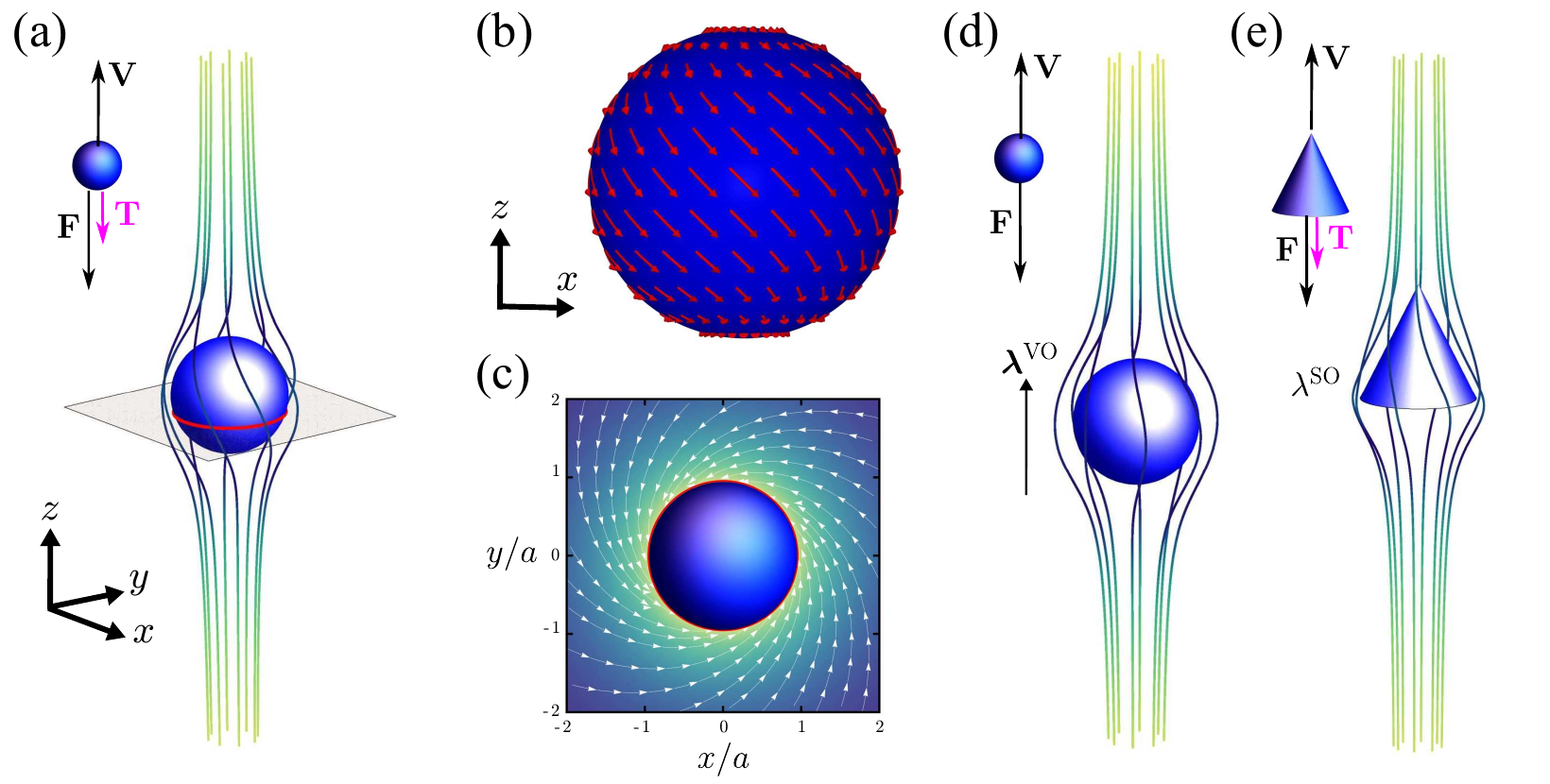}
\caption{
\label{fig:3}
Flow past translating objects with odd surface slip, shown in the co-moving frame.
(a) Flow past a sphere with a scalar odd slip ($\lambda^\text{SO}=0.3a$).
(b) The induced slip velocity on the surface. (c) $x$-$y$ component of the velocity, evaluated at $z=-0.3a$ (gray square in panel (a)).
(d) Flow past a sphere with a vectorial odd slip ($\lambda^\text{VO}=0.3a$).
(e) Flow past a cone with a scalar odd slip ($\lambda^\text{SO}=0.3a$).
}
\end{figure*}

\subsection{Lorentz reciprocal theorem with slip boundary conditions}

We consider the flow around a rigid body with surface $\mathcal{S}$ moving with translational $\mathbf{V}$ and rotational velocities $\boldsymbol{\Omega}$ in an unbounded incompressible Newtonian fluid, which follows the Stokes equations~\eqref{eq:stokes}.
The linear background flow field is given by
$\mathbf{v}^\infty(\mathbf{r})=\mathbf{E}^\infty\cdot\mathbf{r},$ where $\mathbf{E}^\infty$ represents the local strain-rate tensor and is a constant, symmetric, and traceless second-order tensor.
The boundary condition for the velocity field at the surface of the body takes the form
\begin{align}
    \mathbf{v} = 
    \mathbf{V}
    +\boldsymbol{\Omega}\times\mathbf{r}
    +
    \mathbf{v}_\text{S}
     , \quad \mathbf{r}\in\mathcal{S} \label{eq:BC}.
\end{align}
Here, $\mathbf{v}_\text{S}$ is a surface slip velocity in the co-moving frame and satisfies the kinematic (or impermeability) boundary condition, $\mathbf{\hat{n}}\cdot\mathbf{v}_\text{S} =0$.

Let $\mathbf{v}$ and $\boldsymbol{\sigma}$ represent the velocity and the corresponding stress field of the Stokes flow induced by a particle with a slip boundary condition at the surface $\mathcal{S}$. 
The auxiliary flow around a slip particle with the same geometry, denoted by a hat, consists of the velocity $\hat{\mathbf{v}}$ and stress $\hat{\boldsymbol{\sigma}}$.
Noting that the integrals over surfaces at infinity become zero and no body force is present, the Lorentz reciprocal theorem takes the form of~\cite{masoud2019}
\begin{multline}
    \int_\mathcal{S} dS\, 
    (
    \mathbf{v}-\mathbf{E}^\infty\cdot\mathbf{r}
    )
    \cdot
    (
    \hat{\boldsymbol{\sigma}}-
    2\eta
    \hat{\mathbf{E}}^\infty
    )
    \cdot\mathbf{\hat{n}}
    \\    
    =\int_\mathcal{S} dS\, 
    (
    \hat{\mathbf{v}}-\hat{\mathbf{E}}^\infty\cdot\mathbf{r}
    )
    \cdot
    (
    \boldsymbol{\sigma}-
    2\eta
    \mathbf{E}^\infty
    )
   \cdot\mathbf{\hat{n}}
    .
\end{multline}
The quadratic terms in velocity gradients cancel out upon applying the divergence theorem and we obtain
\begin{multline}
    \int_\mathcal{S} dS\, 
    (\mathbf{v}\cdot\hat{\mathbf{f}}
    -\mathbf{E}^\infty:\mathbf{r}\hat{\mathbf{f}}
    -
    2\eta
    \hat{\mathbf{E}}^\infty:\mathbf{v}\mathbf{\hat{n}}
    )
    \\
    =\int_\mathcal{S} dS\, 
    (
    \hat{\mathbf{v}}\cdot\mathbf{f}
    -\hat{\mathbf{E}}^\infty:\mathbf{r}\mathbf{f}
    -
    2\eta
    \mathbf{E}^\infty:\hat{\mathbf{v}}\mathbf{\hat{n}}
    )
    ,
\end{multline}
where the symbol $:$ denotes a twofold contraction $(\mathbf{E}^\infty:\mathbf{v}\mathbf{\hat{n}}=E^\infty_{ij}v_{i}\hat{n}_j)$ and the tractions are defined as $\mathbf{f}=\boldsymbol{\sigma}\cdot \hat{\mathbf{n}}$. 
After applying the boundary conditions~\eqref{eq:BC} on the surface $\mathcal{S}$, we find
\begin{multline}
    \mathbf{V}\cdot\hat{\mathbf{F}}
    +
    \boldsymbol{\Omega}\cdot\hat{\mathbf{T}}
    -
    \mathbf{E}^\infty
    :
    \hat{\mathbf{S}}
    +
    \int_\mathcal{S} dS\, 
    \mathbf{v}_\text{S}\cdot\hat{\mathbf{f}}
   \\
   =
    \hat{\mathbf{V}}\cdot\mathbf{F}
    +
    \hat{\boldsymbol{\Omega}}\cdot\mathbf{T}
    -
    \hat{\mathbf{E}}^\infty
    :
    \mathbf{S}
        +
    \int_\mathcal{S} dS\, 
    \hat{\mathbf{v}}_\text{S}\cdot\mathbf{f}
    ,
    \label{eq:slip_slip}
\end{multline}
where the net force, torque, and the stresslet are given by~\cite{kim2013microhydrodynamics, guazzelli2011physical}
\begin{align}
    \mathbf{F}&=\int_\mathcal{S}dS\,\mathbf{f}, 
    \notag \\
    \mathbf{T}&=\int_\mathcal{S}dS\,\mathbf{r}\times\mathbf{f},
    \notag\\
    \mathbf{S}
    &= 
    \int_\mathcal{S}dS\,
    \left[
    \frac{1}{2}
    \left(
    \mathbf{r}\mathbf{f} 
    +
    \mathbf{f}\mathbf{r}
    \right)
    -
    \frac{1}{3}
    (\mathbf{r}\cdot\mathbf{f})
    \mathbf{I}
    -
    \eta
     (\mathbf{v\hat{n}} + \mathbf{\hat{n}v})
    \right]
    .
    \label{eq:forcemoment}
\end{align}
By introducing the generalized velocities $\bm{\mathsf{V}}= [\mathbf{V}\, ,\boldsymbol{\Omega}\, ,{-\mathbf{E}^\infty}]$ and forces $\bm{\mathsf{F}}= [\mathbf{F}\,,\mathbf{T}\, , \mathbf{S}]$, Eq.~\eqref{eq:slip_slip} can be written in the compact form
\begin{align}
    \bm{\mathsf{V}}\cdot\bm{\hat{\mathsf{F}}}
    +
    \int_\mathcal{S} dS\, 
    \mathbf{v}_\text{S}\cdot\mathbf{\hat{f}}
    =
    \bm{\hat{\mathsf{V}}}\cdot\bm{\mathsf{F}}
    +
    \int_\mathcal{S} dS\, 
    \mathbf{\hat{v}}_\text{S}\cdot\mathbf{f}
    .
    \label{eq:lrt-grand}
\end{align}

\subsection{Onsager-Casimir reciprocity}

The resistance tensors of passive particles are always symmetric ($\bm{\mathsf{R}}^\top=\bm{\mathsf{R}}$) -- a property closely related to Onsager's reciprocity in statistical thermodynamics~\cite{doi2013soft}. With odd slip, this is no longer the case. In the following, we show that a generalized relationship can still be derived. 

We write the slip condition in its most general tensorial form
\begin{align}
    \mathbf{v}_\text{S}=
    \bm{\mathcal{M}}\cdot\mathbf{f}
    ,
    \label{eq:vsMf}
\end{align}
where $\bm{\mathcal{M}}$ represents the coefficient matrix.
If the main problem has the slip tensor $\bm{\mathcal{M}}$ and the auxiliary problem $\hat{\bm{\mathcal{M}}}$, the Lorentz reciprocal theorem as stated in Eq.~\eqref{eq:lrt-grand} leads to the identity
\begin{align}
    \bm{\mathsf{V}}\cdot\hat{\bm{\mathsf{F}}}
    +
    \int_\mathcal{S} dS\, 
    \hat{\mathbf{f}}\cdot\bm{\mathcal{M}}\cdot\mathbf{f}
    =
    \hat{\bm{\mathsf{V}}}\cdot\bm{\mathsf{F}}
    +
    \int_\mathcal{S} dS\, 
    \mathbf{f}
    \cdot\hat{\bm{\mathcal{M}}}\cdot\hat{\mathbf{f}}
    .
\end{align}
After introducing the resistance matrices,  $\bm{\mathsf{F}}=-\bm{\mathsf{R}}\cdot \bm{\mathsf{V}}$  and $\hat{\bm{\mathsf{F}}}=-\hat{\bm{\mathsf{R}}}\cdot \hat{\bm{\mathsf{V}}}$, 
we find
\begin{align}
    {\bm{\mathsf{V}}}\cdot( 
    \bm{\mathsf{R}}^\top-\hat{\bm{\mathsf{R}}})\cdot \hat{\bm{\mathsf{V}}}
    =
    \int_\mathcal{S} dS\,
    \mathbf{f}\cdot
    (
    \bm{\hat{\mathcal{M}}}
    -
    \bm{\mathcal{M}}^\top
    )
    \cdot\hat{\mathbf{f}}
    .
\end{align}
With the choice of
\begin{align}
    \bm{\hat{\mathcal{M}}}
    =
    \bm{\mathcal{M}}^\top
    ,
    \label{eq:conditionM}
\end{align}
the integral on the r.h.s.\ vanishes and we obtain the symmetry relation
\begin{align}
    \bm{\mathsf{R}}^\top=\hat{\bm{\mathsf{R}}}
    . \label{eq:oc1}
\end{align}

For the even and odd slip model introduced in Eq.~\eqref{eq:chiralBC}, Condition~\eqref{eq:conditionM} can be satisfied if the even slip length in the auxiliary problem is identical to that in the main problem, while the odd slip changes its sign, namely
\begin{align}
    \hat{\lambda}^\text{e}=
    \lambda^\text{e},
    \quad
    \hat{\lambda}^\text{o}=
    -\lambda^\text{o}
    ,
\end{align}
and Eq.~\eqref{eq:oc1} can be written as 
\begin{equation}
    \bm{\mathsf{R}}^\top(\lambda^\text{e},\lambda^\text{o})=\bm{\mathsf{R}}(\lambda^\text{e},-\lambda^\text{o})\,.
    \label{eq:symmetryR}
\end{equation}

The anti-symmetric part of the resistance tensor thus changes sign with a change of sign in the odd slip. 
This type of relation is known as Onsager-Casimir reciprocity which holds in a wide range of systems with broken time-reversal symmetry~\cite{fruchart2023odd,everts2024dissipative}.

\subsection{Perturbation solution}
To understand the effect of odd slip on microparticle dynamics, we calculate its leading order contribution to the resistance tensor using a perturbation approach.
The Lorentz reciprocal theorem allows us to determine the slip-induced correction with the knowledge of the exact solutions of the no-slip problem~\cite{ramachandran2009dynamics,masoud2019,arbib2025effective}.
We consider the main problem of a rigid body with shape $\mathcal{S}$ and slip boundary condition and the auxiliary problem of a particle with the same geometry and no-slip (NS) boundary condition.
The reciprocal theorem, Eq.~\eqref{eq:lrt-grand}, connects the two problems through the following relationship
\begin{align}
    \bm{\mathsf{V}}_\text{NS}\cdot\bm{\mathsf{F}}
    =
    \bm{\mathsf{V}}\cdot\bm{\mathsf{F}}_\text{NS}
    +
    \int_\mathcal{S} dS\, 
    \mathbf{v}_\text{S}\cdot\mathbf{f}_\text{NS}
    ,
    \label{eq:LRT}
\end{align}
where we have set $\hat{\mathbf{v}}_\text{S}=\mathbf{0}$.

We now express the forces with the velocities as $\bm{\mathsf{F}}=-\bm{\mathsf{R}}\cdot \bm{\mathsf{V}}$  and ${\bm{\mathsf{F}}}_\text{NS}=-{\bm{\mathsf{R}}}_\text{NS}\cdot {\bm{\mathsf{V}}}_\text{NS}$. The slip velocity is determined by the traction $\mathbf{f}$ through the boundary condition~\eqref{eq:chiralBC}, in which we only keep the odd slip length $\lambda^\text{o}$.
The effect of even slip $\lambda^\text{e}$ can be included through linear superposition, but we do not consider it further as it only contributes to the symmetric parts of the resistance tensor in the perturbative calculation and is not involved in any odd responses discussed here.
Equation~\eqref{eq:LRT} can then be reordered to express the slip-induced corrections to the resistance
\begin{align}
    \bm{\mathsf{V}}_\text{NS}
    \cdot
    (
    \bm{\mathsf{R}}
    -
    \bm{\mathsf{R}}_\text{NS})
    \cdot\bm{\mathsf{V}}
    =
    -
    \frac{1}{\eta}
    \int_\mathcal{S}dS\,
    \lambda^\text{o}(\mathbf{\hat{n}}\times\mathbf{f})\cdot\mathbf{f}_\text{NS}
    \label{eq:VRV}
    .
\end{align}
Because the tractions $\mathbf{f}$  and $\mathbf{f}_\text{NS}$ are linear functions of $\bm{\mathsf V}$ and $\bm{\mathsf{V}}_\text{NS}$, respectively, the integral in Eq.~\eqref{eq:VRV} is a bilinear form of the velocities, which can be eliminated on both sides to formally solve for $\bm{\mathsf{R}}$.
Practically, by choosing the directions of $\bm{\mathsf{V}}_\text{NS}$ and $\bm{\mathsf{V}}$, the sub-matrices in  $\bm{\mathsf{R}}$ can be evaluated individually.

In the limit of small slip lengths compared with the characteristic length of the particle $a$ $(|\lambda^\text{o}|\ll a)$, we can evaluate the integral using a perturbative expansion in the traction~\cite{masoud2019}.
Since the terms on the r.h.s.\ are already linear in the slip length, the traction $\mathbf{f}$ can be expanded around the NS problem up to the zeroth order $\mathbf{f}(\mathbf{V})=\mathbf{f}_\text{NS}(\mathbf{V})+\mathcal{O}(|\lambda^\text{o}|/a)$. With the knowledge of the solution of the no-slip problem, the leading correction to the resistance tensor requires merely the evaluation of a surface integral.

\subsection{Sphere with pseudo-scalar odd slip}
\label{sec:PS}

We first examine the rigid body resistance of a spherical particle of radius $a$ with a pseudo-scalar odd (SO) boundary condition (Fig.~\ref{fig:3}a-c). 
While spherical symmetry forbids any off-diagonal coupling in the resistance tensor~\cite{happel2012low}, chiral slip can break the symmetry of the surface velocity profile, lifting this constraint.
The resistance tensor is obtained to the linear order in $ \lambda^\text{SO}/a$ by evaluating the surface integral in Eq.~\eqref{eq:VRV} using the known solutions for the traction on a sphere (see Appendix~\ref{app:SOVO} for the derivation)
and reads
\begin{align}
    \bm{\mathsf{R}}=
    \bm{\mathsf{R}}_\text{NS}
    +
    (\lambda^\text{SO}/a)
    \begin{pmatrix}
        \mathbf{0} & \bm{\mathcal{R}}  & \mathbf{0} \\
        -\bm{\mathcal{R}}^\top & \mathbf{0} & \mathbf{0} \\
        \mathbf{0} & \mathbf{0} & \mathbf{0} 
    \end{pmatrix}
    ,
\end{align}
with $\bm{\mathcal{R}}=-12\pi\eta a^2\mathbf{I}$.
The resistance satisfies the Onsager-Casimir symmetry formulated in Eq.~\eqref{eq:symmetryR}. 
Despite the symmetric geometry of the sphere, pseudo-scalar slip leads to translation-rotational coupling. For example, a sedimenting sphere now acquires a spinning motion due to the chiral surface activity.

\subsection{Sphere with pseudo-vectorial odd slip}
\label{sec:PV}

We expect that a pseudo-vectorial odd (VO) slip, which breaks the spatial isotropy (Fig.~\ref{fig:3}d), will have a profoundly different effect on the resistance matrix of a sphere. 
We choose the VO slip as $\lambda^\text{o}=\bm\lambda^\text{VO}\cdot\mathbf{\hat{n}}$ with its magnitude $\lambda^\text{VO}$ and externally set direction $\mathbf{\hat{e}}$ and evaluate the integral in Eq.~\eqref{eq:VRV} to $\mathcal{O}(\lambda^\text{VO}/a)$. The resulting resistance tensor has the form (see Appendix~\ref{app:SOVO} for the derivation)
\begin{align}
    \bm{\mathsf{R}}=
    \bm{\mathsf{R}}_\text{NS}
    +
    (\lambda^\text{VO}/a)
    \begin{pmatrix}
        \bm{\mathcal{T}}& \mathbf{0} & \mathbf{0} \\
        \mathbf{0} & \bm{\mathcal{U}} & \bm{\mathcal{X}} \\
        \mathbf{0} & 
        -\bm{\mathcal{X}}^\top & \bm{\mathcal{W}}
    \end{pmatrix}
    ,
    \label{eq:RVO}
\end{align}
where
\begin{align}
    \bm{\mathcal{T}}
    &=
    3\pi\eta a
    \mathbf{\hat{e}}\cdot
    \bm{\epsilon},
    \quad
    \bm{\mathcal{U}}
    =
    12\pi\eta a^3
    \mathbf{\hat{e}}\cdot
    \bm{\epsilon},\notag \\
    \mathcal{X}_{ijk}
    &=
    6\pi\eta a^3
    \hat{e}_q
    \left(
    \delta_{qj}\delta_{ik}
    +
    \delta_{qk}\delta_{ij}
    -
    \frac{2}{3}
    \delta_{qi}
    \delta_{jk}
    \right)
    , \\
    \mathcal{W}_{ijk\ell}
    &=
    \frac{5}{3}
    \pi\eta a^3
    \hat{e}_q
    (
    \epsilon_{qik}\delta_{j\ell}
    +
    \epsilon_{qi\ell}\delta_{jk}
    +
    \epsilon_{qjk}\delta_{i\ell}
    +
    \epsilon_{qj\ell}\delta_{ik}
    )
    .
    \notag 
\end{align}
The matrix $\bm{\mathcal{W}}$ expresses an off-diagonal slip-induced stresslet, which only arises with the odd pseudo-vectorial slip. The induced stresslet is rotated with respect to the applied shear flow (Fig.~\ref{fig:4}a). 
In line with the Onsager-Casimir reciprocity, it is anti-symmetric upon the exchange of indices $ij\leftrightarrow k\ell$.

\subsection{Conical particle with odd pseudo-scalar slip}

\begin{figure*}
\centering
\includegraphics[width=16cm]{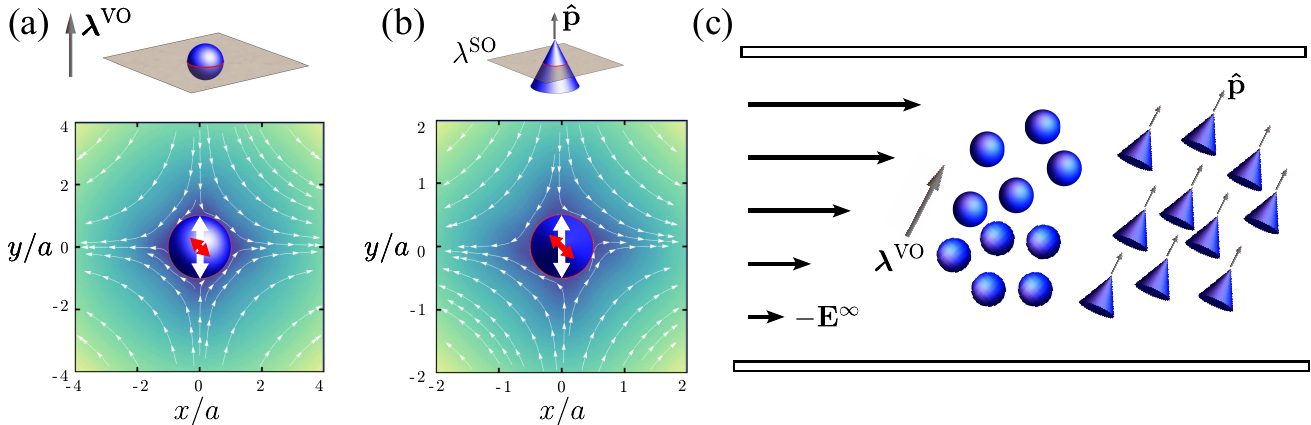}
\caption{
\label{fig:4}
(a) Shear flow past a sphere with a pseudo-vectorial odd slip of magnitude $\lambda^\text{VO}=0.1a$. The induced stresslet moment (white arrows show its principal axis in the absence of odd slip) acquires an odd component in the direction rotated by $+
\pi/4$ (red arrows). 
(b) Shear flow past a cone with a pseudo-scalar odd slip with $\lambda^\text{SO}=0.1a$.
As in panel (a), the flow becomes skewed in the presence of the chiral slip and the induced stresslet is no longer aligned with the applied shear. 
(c)
Both types of objects, i.e. spheres with a pseudo-vectorial odd slip from panel (a) and cones with polar order and pseudo-scalar odd slip from panel (b), exhibit odd viscosity in a suspension. 
}
\end{figure*}

Instead of vectorial orientation dependence of the odd slip, spatial isotropy can also be broken through the asymmetric shape of the object. We demonstrate this by calculating the resistance of a right circular cone with a pseudo-scalar odd slip boundary (Fig.~\ref{fig:3}e). 
In the limit of a small slip length, we find from symmetry considerations that
\begin{align}
    \bm{\mathsf{R}}
    =
    \bm{\mathsf{R}}_\text{NS}
    +
    (\lambda^\text{SO}/a)
    \begin{pmatrix}
        \bm{\mathcal{T}}
        & \bm{\mathcal{R}} & \bm{\mathcal{V}}\\
        -\bm{\mathcal{R}}^\top & \bm{\mathcal{U}}& \bm{\mathcal{X}} \\
        -\bm{\mathcal{V}}^\top & 
        -\bm{\mathcal{X}}^\top & \bm{\mathcal{W}}
        \label{eq:RSOcone}
    \end{pmatrix}
    ,
\end{align}
where
\begin{align}
    \bm{\mathcal{T}}
    &=
    \mathcal{T}
    \mathbf{\hat{p}}
    \cdot
    \bm\epsilon
    ,
    \quad
    \bm{\mathcal{R}}
    =
    \mathcal{R}_\|
    \mathbf{\hat{p}\hat{p}}
    +
    \mathcal{R}_\perp
    \mathbf{P}^\perp
    ,
    \quad
    \bm{\mathcal{U}}
    =
    \mathcal{U}
    \mathbf{\hat{p}}
    \cdot
    \bm\epsilon
    ,
    \nonumber
    \\
    \mathcal{V}_{ijk}
    &=
    \mathcal{V}
    \hat{p}_\ell
    (
    \epsilon_{ij\ell}\hat{p}_k
    +
    \epsilon_{ik\ell}\hat{p}_j
    )
    \nonumber\\
    \mathcal{X}_{ijk}
    &=
    \mathcal{X}_\|
    \hat{p}_i
    \left(
    \hat{p}_j\hat{p}_k
    -
    \frac 13 \delta_{jk}
    \right)
    +
    \mathcal{X}_\perp
    \left(
    P^\perp_{ij}
    \hat{p}_k
    +
    P^\perp_{ik}
    \hat{p}_j
    \right)
    ,
    \label{eq:RSOconesub}\\
    \mathcal{W}_{ijk\ell}
    &=
    \mathcal{W}_\perp
    \hat{p}_q
    \left(
    \epsilon_{qik}
    P^\perp_{\ell j}
    +
    \epsilon_{qi\ell}
    P^\perp_{kj}
    +
    \epsilon_{qjk}
    P^\perp_{\ell i}
    +
    \epsilon_{qj\ell}
    P^\perp_{ki}
    \right)
    \nonumber\\
    &-
    \mathcal{W}_\|
    \hat{p}_q
    \left(
    \epsilon_{qik}
    \hat{p}_\ell \hat{p}_j
    +
    \epsilon_{qi\ell}
    \hat{p}_k \hat{p}_j
    +
    \epsilon_{qjk}
    \hat{p}_\ell \hat{p}_i
    +
    \epsilon_{qj\ell}
    \hat{p}_k \hat{p}_i
    \right)
    \nonumber
    ,
\end{align}
where $a$ is the radius of the cone, $\mathbf{\hat{p}}$ a unit vector along its symmetry axis, and $\mathbf{P}^\perp=\mathbf{I}-\mathbf{\hat{p}\hat{p}}$ a projection operator to the plane normal to the axis.
The combination of the shape asymmetry (determined by the vector $\hat{\mathbf{p}}$) and a pseudo-scalar odd-slip surface indeed results in a resistance tensor with a similar structure to the sphere with pseudo-vectorial odd slip (Fig.~\ref{fig:4}b). The numerical values of the coefficients, computed using the boundary element method (BEM), are shown in Appendix~\ref{app:cone}.

\section{Emergence of odd viscosity in a suspension}

In the following, we examine how the anomalous transport coefficients of a particle affect the rheological properties of a metafluid consisting of a suspension of such particles. 
The macroscopic stress of a homogeneous suspension is given by~\cite{batchelor1970stress}
\begin{align}
 \mathbf{\Sigma}
    =
    -
    \langle p\rangle
    \mathbf{I}
    +
    2\eta \langle\mathbf{E}\rangle
    +
    \mathbf{\Sigma}^\text{p},
    \label{eq:suspension}
\end{align}
where $\langle p\rangle$ and $\langle\mathbf{E}\rangle$ are the mean pressure and rate of strain and $\mathbf{\Sigma}^\text{p}$ represents the stress contribution from the suspended particles. The excess stress due to particles contains two contributions: the mean stresslet and the mean torque from particles.
For a sufficiently dilute suspension, hydrodynamic interactions between the particles can be neglected and the particle stress can be evaluated by the force moments of a single particle in a shear flow.
If we denote by $n$ the number density of particles, their contribution to the stress becomes linear in $n$, namely, $\boldsymbol{\Sigma}^\text{p}=n\left(\mathbf{S}-\frac{1}{2}\boldsymbol{\epsilon}\cdot\mathbf{T}\right)$.
Because the suspended particles are force- and torque-free, the stresslet on a particle is obtained by solving the matrix equation $[\bm{0},\bm{0},\mathbf{S}]= -\bm{\mathsf{R}} \cdot [\mathbf{V},\boldsymbol\Omega,-\mathbf{E}^\infty]$. 
For a sphere with pseudo-vectorial slip and the resistance tensor given by Eq.~\eqref{eq:RVO}, the resulting stresslet moment is $\mathbf{S}=(\mathbf{R}^{SE}-\mathbf{R}^{S\Omega}\cdot(\mathbf{R}^{T\Omega})^{-1}\cdot\mathbf{R}^{TE}):\mathbf{E}^\infty$.
However, the second term only contributes to quadratic order in the slip length.

Writing the constitutive relation $\mathbf{\Sigma}=-p\mathbf{I}+\boldsymbol{\eta}^\text{eff}:\nabla\mathbf{v}^\infty$, the viscosity tensor of the suspension of spheres with vectorial slip evaluates to 
\begin{align}
    &\eta^\text{eff}_{ijk\ell}
    =
    \eta \left(1+\frac{5}{2}\phi\right)
    \left(
    \delta_{ik}\delta_{j\ell}
    +
    \delta_{i\ell}\delta_{jk}
    -
    \frac{2}{3}
    \delta_{ij}\delta_{k\ell}
    \right)
    \nonumber\\
    &+
    \frac{5}{4}\eta\phi
    \frac{\lambda^\text{VO}_q}{a}
    \left(
    \epsilon_{qik}\delta_{j \ell}
    +
    \epsilon_{qi\ell}\delta_{jk}
    +
    \epsilon_{qjk}\delta_{i\ell}
    +
    \epsilon_{qj\ell}\delta_{ik}
    \right)
    ,
    \label{eq:eta_ijkl_vo}
\end{align}
with $\phi=4\pi a^3n/3$ being the volume fraction of the spheres.
The first term represents the well-known Einstein viscosity of a suspension of no-slip spheres.
The second term corresponds exactly to the odd viscosity as defined, e.g., in Refs.~\cite{markovich2021, khain2022, hosaka2023lorentz, messica2025stokes} (see Ref.~\cite{hosaka2023lorentz} for conversion factors), with 
\begin{equation}\eta^\text{o}= \frac{5 \eta \phi \lambda^\text{VO}}{2a}\,.\end{equation} 
Our results show that the pseudo-vectorial odd slip on the surface of otherwise symmetric particles can lead to an odd viscosity (itself an axial vector) of the suspension (Fig.~\ref{fig:4}c).

Using the typical values $\phi\approx0.1$, a cilia length $L\approx 0.1\, a$ and the attainable odd slip $\lambda^{\rm o}\approx L$, the odd viscosity is of the order  $\eta^{\rm o}\approx 0.025\,\eta$, sufficient to produce distinct anti-symmetric responses.

Likewise, a suspension of cones with pseudo-scalar odd slip and polar alignment also shows odd viscosity (Appendix~\ref{app:cone}). However, instead of the simple form of Eq.~\eqref{eq:eta_ijkl_vo}, the tensor contains two independent odd viscosities, which are generically allowed in a system with cylindrical symmetry \cite{khain2022,MeissnerOszer.Everts2026}. The structure of the viscosity tensor is similar to that of a suspension of no-slip particles in a fluid with intrinsic odd viscosity \cite{everts2024dissipative}. 
In addition, there are odd transport coefficients coupling shear and rotation.
Furthermore, asymmetry in the odd-slip distribution can also lead to odd viscosity in suspensions.
One of the realizations is an odd-slip Janus sphere, a particle with pseudo-scalar odd slip surface on one half and no-slip on the other (see Appendix~\ref{app:Janus} for the solution).
From the symmetry perspective, odd viscosity, which requires broken parity and spatial anisotropy, can either be achieved with a single axial vector (namely, the vectorial odd slip), or the combination of a pseudo-scalar (odd slip) and a polar vector, describing the particle alignment.

We summarize that just like the broken parity symmetry in a fluid can lead to an asymmetric particle mobility \cite{khain2024trading}, particles with a parity-violating surface slip can form a metafluid with odd macroscopic viscosity.

\section{Conclusion and Outlook}

Odd slip is a parity- and time-reversal symmetry breaking response on active surfaces interfacing with three-dimensional fluids. In terms of dimensionality, it bridges the gap between the odd diffusivity of point particles and odd viscosity or elasticity of bulk media. We showed that flexible cilia, which are actively driven in a chiral way and therefore do break both the time-reversal and the parity symmetry, lead to an effective odd slip on the surface when viewed in a coarse-grained way. Even though most cilia do not beat in a symmetric way assumed in our model, we expect that they can still exhibit odd slip in superposition to the active flows they generate. 
Beyond the present dilute ciliary model, related odd-slip responses may also arise from
hydrodynamic interactions among densely packed cilia or flagella that can display metachronal
waves, nonlinear synchronization, and shear-induced transitions between collective states~\cite{Osterman.Vilfan2011,martindale2017autonomously}. The microscopic model allowed us to calculate the amount of dissipation needed to produce a slip that appears dissipationless in the macroscopic picture. We expect that similar relations can be found for other non-dissipative phenomena including odd viscosity. 

Other possible realizations of odd slip include a carpet of spinning colloidal magnets~\cite{soni2019odd} or any other mechanism that produces a thin layer of an odd-viscous fluid. A recent preprint shows that a related odd response can be achieved in a carpet of rotating bacterial flagella \cite{chahal2026emergentoddresponseactive}.
Furthermore, odd slip can theoretically arise in a charged boundary layer subjected to a magnetic field. The latter possibility is notable because it shows that a magnetic field can be sufficient to provide the broken time-reversal and parity symmetry and that odd slip does not necessarily need activity with constant energy input. However, the expected effect is weak and, like in other colloidal systems, activity is a more promising way towards odd systems.

The odd slip leads to intriguing hydrodynamic problems, of which we have solved only the most basic examples. These have demonstrated how the broken symmetries in the boundary condition together with the broken symmetries of the body shape lead to parity-violating terms in the mobility. We also demonstrated an intriguing interconnectedness between ``odd'' phenomena in different dimensions: a suspension of particles with odd slip and broken spatial isotropy creates a metafluid that exhibits odd viscosity. 
Experimental realizations of odd viscosity in 3D colloidal systems are still elusive and even the microscopic mechanisms of its emergence or its minimum requirements in fluidic environments are still incompletely understood \cite{hargus2020time, markovich2021, han2021fluctuating, eren2026hydrodynamic}. 
Therefore, active surfaces could provide an entirely different approach towards the realization of metafluids with odd viscosity.

\begin{acknowledgments}
Y.H.\ acknowledges support from the Japan Society for the Promotion of Science (JSPS) Overseas Research Fellowships (Grant No.\ 202460086) and the Japan Science and Technology Agency (JST) CREST (Grant No.\ JPMJCR25Q1).
A.V.\ acknowledges support from the Slovenian Research and Innovation Agency (Grants No.\ P1-0099 and J1-60009).
\end{acknowledgments}

\begin{widetext}
\appendix

\section{Perturbative calculation of the grand resistance tensor of an odd-slip sphere}
\label{app:SOVO}

\subsection{Decomposition of the traction fields}

The traction $\mathbf{f}=\boldsymbol\sigma \cdot \hat{\mathbf{n}}$ describes the position-dependent force density exerted by the fluid on the surface of a particle. For a particle moving with velocity $\mathbf{V}$ and angular velocity $\boldsymbol\Omega$ in a flow with shear rate $\mathbf{E}^\infty$, their contributions can be decomposed such that~\cite{pozrikidis1992}
\begin{align}
    \mathbf{f} =
    \eta(
    \mathbf{d}^V\cdot\mathbf{V} 
    +
    \mathbf{d}^\Omega\cdot\bm{\Omega} 
    +
    \mathbf{d}^E:\mathbf{E}^\infty 
    ).
    \label{eq:d-decomposition}
\end{align}
For a no-slip sphere of radius $a$, the solutions are given by~\cite{happel2012low}
\begin{align}
    d^V_{ij}=-3/(2a)\delta_{ij},\quad
    d_{ij}^\Omega=-3\epsilon_{ijk}\hat{n}_k,\quad
    d^E_{ijk}=
    5(\delta_{ij}\hat{n}_k+\delta_{ik}\hat{n}_j
    -(2/3) \hat{n}_i\delta_{jk}
    )/2
    .
\end{align}

Inserting these into the force moments~\eqref{eq:forcemoment} yields the well-known results for a spherical body with no-slip boundaries.
The corresponding resistance functions are given by
\begin{align}
    R_{\text{NS},ij}^{ FV}&= 6\pi\eta a \delta_{ij}, 
    &
    R_{\text{NS},ij}^{  F\Omega}&= 
    0
    , 
    &
    R_{\text{NS},ijk}^{FE}&= 0,
    \\
    R_{\text{NS},ij}^{ TV} &= 
    0,
    &
    R_{\text{NS},ij}^{ T\Omega} &= 8\pi\eta a^3\delta_{ij},
    &
    R_{\text{NS},ijk}^{ TE} &= 0,
    \\ 
    R_{\text{NS},ijk}^{  SV} &= 0,
    &
    R_{\text{NS},ijk}^{ S\Omega}&= 0,
    &
    R_{\text{NS},ijk\ell}^{ SE}&= 
    \frac{10}{3}\pi\eta a^3
    \left(
    \delta_{ik}\delta_{j\ell}
    +
    \delta_{i\ell}\delta_{jk}
    -\frac{2}{3}
    \delta_{ij}\delta_{k\ell}
    \right)
    .
\end{align}

\subsection{Pseudo-scalar (scalar odd: SO) slip sphere}

Here we determine the leading-order effect of the pseudo-scalar slip on the grand resistance tensor of a rigid sphere.
The resistance tensor can be expressed with the integrals of the traction fields using Eq.~\eqref{eq:VRV} from the main text
\begin{align}
    \bm{\mathsf{V}}_\text{NS}
    \cdot
    \bm{\mathsf{R}}
    \cdot
    \bm{\mathsf{V}}
    =
    \bm{\mathsf{V}}
    \cdot
    \bm{\mathsf{R}}_\text{NS}
    \cdot
    \bm{\mathsf{V}}_\text{NS}
    -
    \frac{\lambda^\text{SO}}{\eta}
    \int_\mathcal{S}dS\,
    (\mathbf{\hat{n}}\times\mathbf{f})\cdot\mathbf{f}_\text{NS}
    .
    \label{eq:LRT_oPS}
\end{align}
To obtain the first-order correction in $\lambda^\text{SO}/a$, the unknown traction field can be expanded up to the zeroth order around the auxiliary problem, namely, using the NS solutions themselves. 
From the symmetry of the sphere, it follows that the only non-vanishing first-order slip-induced corrections are the velocity-induced torque matrix and its transposed
counterpart.
By choosing a translating sphere with $\mathbf{V}$ as the main problem and a rotating sphere with $\boldsymbol{\Omega}_\text{NS}$ as the auxiliary problem, the integral in Eq.~\eqref{eq:LRT_oPS}, to leading order, evaluates to
\begin{align}
    \boldsymbol{\Omega}_\text{NS}\cdot\mathbf{R}^{TV} 
    \cdot\mathbf{V} 
    &=
    -\eta\lambda^\text{SO}
    \int_\mathcal{S}dS\,
    [\mathbf{\hat{n}}\times
    (\mathbf{d}^V\cdot\mathbf{V} )
    ]
    \cdot
    (
    \mathbf{d}^\Omega
    \cdot
    \boldsymbol{\Omega}_\text{NS}
    )
    .
\end{align}
Here, we have expressed the tractions with velocities using the functions introduced in Eq.~\eqref{eq:d-decomposition}.
Eliminating the arbitrary velocity vectors isolates the resistance tensor of interest
\begin{align}
    R^{TV}_{ij}
    =
    -\eta\lambda^\text{SO}
    \int_\mathcal{S}dS\,
    \epsilon_{k\ell m}\hat{n}_\ell
    d^{V}_{mj}d^{\Omega}_{ki}
    .
\end{align}
Integrating over the sphere surface, we then find to the linear order in the slip length 
\begin{align}
    R^{TV}_{ij}
    =
    12\pi\eta a^2 (\lambda^\text{SO}/a)
    \delta_{ij},
\end{align}
where the relation $\int_\mathcal{S}dS\, (\mathbf{\hat{n}\hat{n}}-\mathbf{I})=-8\pi a^2/3\mathbf{I}$ has been used.
Similarly, we find 
\begin{align}
    R^{F\Omega}_{ij}
    =
    -12\pi\eta a^2 (\lambda^\text{SO}/a)
    \delta_{ij}
    .
\end{align}
The first-order resistance tensors of a pseudo-scalar odd slip sphere can be summarized as
\begin{align}
    R_{ij}^{ FV}&= R_{\text{NS},ij}^{FV}, 
    &
    R_{ij}^{  F\Omega}&= 
    -12\pi\eta a \lambda^\text{SO}
    \delta_{ij}
    , 
    &
    R_{ijk}^{  FE}&= 0,
    \\
    R_{ij}^{ TV} &= 12\pi\eta a \lambda^\text{SO}
    \delta_{ij},
    &
    R_{ij}^{ T\Omega} &= R_{\text{NS},ij}^{T\Omega},
    &
    R_{ijk}^{ TE} &= 0,
    \\ 
    R_{ijk}^{  SV} &= 0,
    &
    R_{ijk}^{ S\Omega}&= 0,
    &
    R_{ijk\ell}^{ SE}&= 
    R_{\text{NS},ijk\ell}^{SE}
    .
\end{align}
Odd slip results in translational-rotational coupling, but does not have an effect on any other coefficients.

\subsection{Pseudo-vectorial (vectorial odd: VO) slip sphere}

We now determine the leading-order effect of the pseudo-vectorial slip on the grand resistance tensor of a sphere.
From the Lorentz reciprocal theorem, Eq.~\eqref{eq:VRV} in the main text, the resistance matrix follows
\begin{align}
    \bm{\mathsf{V}}_\text{NS}
    \cdot
    \bm{\mathsf{R}} 
    \cdot
    \bm{\mathsf{V}} 
    =
    \bm{\mathsf{V}} 
    \cdot
    \bm{\mathsf{R}}_\text{NS}
    \cdot
    \bm{\mathsf{V}}_\text{NS}
    -
    \frac{1}{\eta}\boldsymbol\lambda^\text{VO}\cdot
    \int_\mathcal{S}dS\,
    \mathbf{\hat{n}}(\mathbf{\hat{n}}\times\mathbf{f} )\cdot\mathbf{f}_\text{NS}
    .
\end{align}
To evaluate first the shear-induced resistance coefficients, we consider the main problem with a free sphere in an imposed flow $\mathbf{E}^\infty \cdot\mathbf{r}$ and the auxiliary one with a NS sphere of the same instantaneous shape in a background flow $\mathbf{E}^\infty_\text{NS}\cdot\mathbf{r}$.
The shear-translational coupling vanishes for symmetry reasons, while the terms linking shear rate with stress and torque remain.
To first order in $\lambda^\text{VO}$, the corrections read
\begin{align}
    \mathbf{E}^\infty_\text{NS}:\mathbf{R}^{SE} 
    :\mathbf{E}^\infty 
    =
    \mathbf{E}^\infty:\mathbf{R}^{SE}_\text{NS} 
    :\mathbf{E}^\infty_\text{NS}
    -
    \eta
    \boldsymbol{\lambda}^\text{VO}
    \cdot
    \int_\mathcal{S}dS\,
    \mathbf{\hat{n}}
    [\mathbf{\hat{n}}\times
    (\mathbf{d}^E:\mathbf{E}^\infty )
    ]
    \cdot
    (
    \mathbf{d}^E
    :
    \mathbf{E}^\infty_\text{NS}
    )
\end{align}
and
\begin{align}
    \boldsymbol{\Omega}_\text{NS}\cdot\mathbf{R}^{TE} 
    :\mathbf{E}^\infty 
    =
    \eta
    \boldsymbol{\lambda}^\text{VO}
    \cdot
    \int_\mathcal{S}dS\,
    \mathbf{\hat{n}}
    [\mathbf{\hat{n}}\times
    (\mathbf{d}^E:\mathbf{E}^\infty )
    ]
    \cdot
    (
    \mathbf{d}^\Omega
    \cdot
    \boldsymbol{\Omega}_\text{NS}
    )
    .
\end{align}
Eliminating the arbitrary vectors and tensors, each resistance matrix can be solved separately to obtain
\begin{align}
    R_{ijk\ell}^{SE}
    =
    R_{\text{NS},ijk\ell}^{SE}
    -
    \eta
    \lambda^\text{VO}_q
    \int_\mathcal{S}dS\,
    \hat{n}_q
    \epsilon_{mnp}\hat{n}_n
    d^E_{pk\ell}
    d^E_{mij}
    ,
\end{align}
and
\begin{align}
    R_{ijk}^{TE}
    =
    \eta
    \lambda^\text{VO}_q
    \int_\mathcal{S}dS\,
    \hat{n}_q
    \epsilon_{mnp}\hat{n}_n
    d^E_{pjk}
    d^\Omega_{mi}
    .
\end{align}
The integrals can be evaluated to give
\begin{align}
    R_{ijk\ell}^{SE}
    =
    R_{\text{NS},ijk\ell}^{SE}
    +
    \frac{5}{3}
    \pi\eta a^3
    (\lambda^\text{VO}_q/a)
    (
    \epsilon_{qik}\delta_{\ell j}
    +
    \epsilon_{qi\ell}\delta_{kj}
    +
    \epsilon_{qjk}\delta_{\ell i}
    +
    \epsilon_{qj\ell}\delta_{ki}
    )
    ,
\end{align}
and
\begin{align}
    R_{ijk}^{TE}
    =
    6\pi\eta a^3
    (\lambda^\text{VO}_q/a)
    \left(
    \delta_{qj}\delta_{ik}
    +
    \delta_{qk}\delta_{ij}
    -
    \frac{2}{3}
    \delta_{qi}
    \delta_{jk}
    \right)
    ,
\end{align}
where we have used the surface integral of the $\mathbf{\hat{n}}$-moments
\begin{align}
    \int_\mathcal{S}dS\, 
    \hat{n}_i\hat{n}_j\hat{n}_k\hat{n}_\ell
    =
    \frac{4}{15}\pi a^2
    (
    \delta_{ij}\delta_{k\ell}
    +
    \delta_{ik}\delta_{j\ell}
    +
    \delta_{i\ell}\delta_{jk}
    )
    .
\end{align}

Similarly, the $\mathbf{V}$- and $\bm\Omega$-induced responses can be calculated from the integral expressions:
\begin{align}
    \mathbf{V}_\text{NS}\cdot\mathbf{R}^{FV} 
    \cdot\mathbf{V} 
    &=
    \mathbf{V}\cdot\mathbf{R}^{FV}_\text{NS} 
    \cdot\mathbf{V}_\text{NS}
    -
    \eta
    \boldsymbol{\lambda}^\text{VO}
    \cdot
    \int_\mathcal{S}dS\,
    \mathbf{\hat{n}}
    [\mathbf{\hat{n}}\times
    (\mathbf{d}^V\cdot\mathbf{V})
    ]
    \cdot
    (
    \mathbf{d}^V
    \cdot
    \mathbf{V}_\text{NS}
    )
    ,
    \\
    \bm\Omega_\text{NS}\cdot\mathbf{R}^{T\Omega} 
    \cdot\bm\Omega
    &=
    \bm\Omega\cdot\mathbf{R}^{T\Omega}_\text{NS} 
    \cdot\bm\Omega_\text{NS}
    -
    \eta
    \boldsymbol{\lambda}^\text{VO}
    \cdot
    \int_\mathcal{S}dS\,
    \mathbf{\hat{n}}
    [\mathbf{\hat{n}}\times
    (\mathbf{d}^\Omega\cdot\bm\Omega)
    ]
    \cdot
    (
    \mathbf{d}^\Omega
    \cdot
    \bm\Omega_\text{NS}
    )
    ,
\end{align}
from which the first-order resistance functions can be isolated:
\begin{align}
    R_{ij }^{FV}
    &=
    R_{\text{NS},ij }^{FV}
    -
    \eta
    \lambda^\text{VO}_q
    \int_\mathcal{S}dS\,
    \hat{n}_q
    \epsilon_{mnp}\hat{n}_n
    d^V_{ pj}
    d^V_{ mi}
    ,\\
    R_{ij }^{T\Omega}
    &=
    R_{\text{NS},ij }^{T\Omega}
    -
    \eta
    \lambda^\text{VO}_q
    \int_\mathcal{S}dS\,
    \hat{n}_q
    \epsilon_{mnp}\hat{n}_n
    d^\Omega_{ pj}
    d^\Omega_{ mi}
    .
\end{align}
Integrating over the sphere surface provides the diagonal resistance matrices
\begin{align}
    R_{ij}^{FV}
    &=
    R_{\text{NS}, ij}^{FV}
    +
    3\pi\eta a
    (\lambda^\text{VO}_q/a)
    \epsilon_{qij}
    ,
    \\
    R_{ij}^{T\Omega}
    &=
    R_{\text{NS}, ij}^{T\Omega}
    +
    12\pi\eta a^3
    (\lambda^\text{VO}_q/a)
    \epsilon_{qij}
    .
\end{align}

In contrast to the pseudo-scalar case, the translational-rotational coupling vanishes.
Summing up the results above, the first-order resistance tensors of a VO-slip sphere are then given by
\begin{align}
    R_{ij}^{ FV}&= 
    R_{\text{NS},ij}^{ FV}
    +
    3\pi\eta \lambda^\text{VO}_q
    \epsilon_{qij}
    , 
    &
    R_{ij}^{ F\Omega}&= 
    0
    , 
    &
    R_{ijk}^{  FE}&= 0, 
    \\
    R_{ij}^{ TV} &= 0,
    &
    R_{ij}^{ T\Omega} &= 
    R_{\text{NS},ij}^{ T\Omega}
    +
    12\pi\eta a^2\lambda^\text{VO}_q
    \epsilon_{qij}
    ,
    &
    R_{ijk}^{  TE} &= 
    6\pi\eta a^2
    \lambda^\text{VO}_q
    \left(
    \delta_{qj}\delta_{ik}
    +
    \delta_{qk}\delta_{ij}
    -
    \frac{2}{3}
    \delta_{qi}
    \delta_{jk}
    \right)
    ,
\end{align}
and
\begin{align}
    R_{ijk}^{ SV} &= 0,\\
    R_{ijk}^{ S\Omega}&=
    -6\pi\eta a^2
    \lambda^\text{VO}_q
    \left(
    \delta_{qj}\delta_{ik}
    +
    \delta_{qi}\delta_{jk}
    -
    \frac{2}{3}
    \delta_{qk}
    \delta_{ij}
    \right)
    ,
    \\
    R_{ijk\ell}^{ SE}&= 
    R_{\text{NS},ijk\ell}^{ SE}
    +
    \frac{5}{3}
    \pi\eta a^2
    \lambda^\text{VO}_q
    (
    \epsilon_{qik}\delta_{\ell j}
    +
    \epsilon_{qi\ell}\delta_{kj}
    +
    \epsilon_{qjk}\delta_{\ell i}
    +
    \epsilon_{qj\ell}\delta_{ki}
    )
    .
\end{align}

\section{Suspension of cones with pseudo-scalar odd slip}
\label{app:cone}

\subsection{
Matrix notation of tensorial structures in irreducible representations of SO(3)}

Higher-rank tensors connecting the stress to the velocity gradient tensors $\mathbf{u}=\nabla\mathbf{v}$ can be expressed in matrix notation by reducing dimensionality with irreducible representations of $\text{SO}(3)$~\cite{scheibner2020odd}.
Apart from the dilatational mode, irrelevant to incompressible fluids we consider, there are eight basis matrices, given by 
\begin{align}
    \boldsymbol{\tau}^{\omega_x}  = 
    \frac{1}{\sqrt{2}}
    \begin{pmatrix}
        0 & 0 & 0\\
        0 & 0 & 1 \\
        0 & -1 & 0\\
    \end{pmatrix}
    ,\quad
    \boldsymbol{\tau}^{\omega_y}  = 
    \frac{1}{\sqrt{2}}
    \begin{pmatrix}
        0 & 0 & -1\\
        0 & 0 & 0 \\
        1 & 0 & 0\\
    \end{pmatrix},
    \quad
    \boldsymbol{\tau}^{\omega_z}  =
    \frac{1}{\sqrt{2}}
    \begin{pmatrix}
        0 & 1 & 0\\
        -1 & 0 & 0 \\
        0 & 0 & 0\\
    \end{pmatrix}
    ,
    \label{eq:basis1}
\end{align}
for three pure rotations, and 
\begin{align}
    \boldsymbol{\tau}^{1} &= 
    \frac{1}{\sqrt{2}}
    \begin{pmatrix}
        1 & 0 & 0\\
        0 & -1 & 0 \\
        0 & 0 & 0\\
    \end{pmatrix}
    ,\quad
    \boldsymbol{\tau}^{2}  =
    \frac{1}{\sqrt{2}}
    \begin{pmatrix}
        0 & 1 & 0\\
        1 & 0 & 0 \\
        0 & 0 & 0\\
    \end{pmatrix}
    ,\quad
    \boldsymbol{\tau}^{3}  = 
    \frac{1}{\sqrt{6}}
    \begin{pmatrix}
        -1 & 0 & 0\\
        0 & -1 & 0 \\
        0 & 0 & 2\\
    \end{pmatrix}
    ,\nonumber\\
    &\boldsymbol{\tau}^{4}  = 
    \frac{1}{\sqrt{2}}
    \begin{pmatrix}
        0 & 0 & 0\\
        0 & 0 & 1 \\
        0 & 1 & 0\\
    \end{pmatrix}
    ,\quad
    \boldsymbol{\tau}^{5}  = 
    \frac{1}{\sqrt{2}}
    \begin{pmatrix}
        0 & 0 & 1\\
        0 & 0 & 0 \\
        1 & 0 & 0\\
    \end{pmatrix}
    ,
    \label{eq:tau1-5}
\end{align}
for five independent shear modes including elongation, pure, and uniaxial strain.
The basis vectors satisfy the orthonormality relation $\tau^\alpha_{ij}\tau^\beta_{ij}=\delta^{\alpha\beta}$ with the two-fold contraction as a scalar product.

As an example, we express the generic viscosity tensor in matrix notation.
The constitutive relation, $\sigma_{ij}=\eta_{ijk\ell}\partial_\ell v_k$, can be expressed in this basis, such that $\tilde{\sigma}^\alpha = \tilde{\eta}^{\alpha\beta}\tilde{u}^\beta$, where the velocity gradient and stress tensors are represented as vectors and the viscosity tensor reduces to a matrix
\begin{align}
    \tilde{u}^\alpha=u_{ij}\tau^\alpha_{ij},\quad
    \tilde{\sigma}^\alpha=\sigma_{ij}\tau^\alpha_{ij},\quad
    \tilde{\eta}^{\alpha\beta}=\tau_{ij}^\alpha\eta_{ijk\ell}\tau^\beta_{k\ell},
\end{align}
with the transformed quantities denoted by a tilde symbol.
The inverse transform is defined as
\begin{align}
    u_{ij} = \tilde{u}^\alpha \tau_{ij}^\alpha,\quad
    \sigma_{ij} = \tilde{\sigma}^\alpha \tau_{ij}^\alpha,\quad
    \eta_{ijk\ell}
    =     \tau^\alpha_{ij}\tilde{\eta}^{\alpha\beta}\tau^\beta_{k\ell}
    .
\end{align}
Similarly, rank-3 tensors can also be expressed in the $\bm\tau$-basis.
For example, the force-strain and stresslet-rotational resistance tensors can be transformed into $3\times5$ and $5\times3$ matrices, respectively, because they only require the traceless and symmetric basis vectors.
Specifically, one can find for $\alpha=1,\dots5$
\begin{align}
    \tilde{R}^{FE,\alpha}_{i}
    =
    R^{FE}_{ijk}\tau^\alpha_{jk},
    \quad
    \tilde{R}^{S\Omega,\alpha}_{k}
    =
    \tau^\alpha_{ij}R^{S\Omega}_{ijk}
    .
\end{align}

\subsection{Generic expressions of the grand resistance tensor satisfying the cylindrical symmetry}

We provide the most general rank 2, 3, and 4 tensors satisfying the cylindrical symmetry about an arbitrary direction, and give their explicit matrix representations used later along with numerical results.
For 3\textsuperscript{rd}- and 4\textsuperscript{th}-rank tensors we consider a tensor that is invariant under the exchange of its second-to-last and last indices and that is traceless when these indices are contracted.
These properties are relevant to the stresslet-strain tensor $\mathbf{R}^{SE}$ and the rank 3 coupling tensors.

We first consider a 2\textsuperscript{nd}-rank resistance tensor satisfying the cylindrical symmetry around a direction $\mathbf{\hat{e}}$.
Under this constraint, the sub-matrices are allowed to take the following forms, each of which is characterized by three coefficients~\cite{khain2024trading}
\begin{align}
    \mathbf{R}^{FV}
    &= 
    A^\text{t} (\mathbf{I}-\mathbf{\hat{e}\hat{e}})
    +
    A^\text{n}\mathbf{\hat{e}\hat{e}}
    +
    A^\text{o}\bm{\epsilon}\cdot\mathbf{\hat{e}},
    &
    \mathbf{R}^{T\Omega}
    &= 
    B^\text{t} (\mathbf{I}-\mathbf{\hat{e}\hat{e}})
    +
    B^\text{n}\mathbf{\hat{e}\hat{e}}
    +
    B^\text{o}\bm{\epsilon}\cdot\mathbf{\hat{e}},
    &
    \\
    \mathbf{R}^{F\Omega}
    &= 
    C^\text{t} (\mathbf{I}-\mathbf{\hat{e}\hat{e}})
    +
    C^\text{n}\mathbf{\hat{e}\hat{e}}
    +
    C^\text{o}\bm{\epsilon}\cdot\mathbf{\hat{e}},
    &
    \mathbf{R}^{TV}
    &= 
    -
    C^\text{t} (\mathbf{I}-\mathbf{\hat{e}\hat{e}})
    -
    C^\text{n}\mathbf{\hat{e}\hat{e}}
    -
    C^\text{o}\bm{\epsilon}\cdot\mathbf{\hat{e}},
    &
    \label{eq:RFVRTV}
\end{align}
where we have used the Onsager-Casimir reciprocity, Eq.~\eqref{eq:symmetryR} in the main text, to reduce the number of independent coefficients.
Note that we have only kept the parity-odd terms, $C^{\rm t}$ and $C^{\rm n}$, in the symmetric parts of the coupling tensors, which are directly relevant to the odd-slip corrections discussed later.

Next, we consider 3\textsuperscript{rd}-rank tensors with the cylindrical symmetry.
As before, they can be expressed in terms of three coefficients~\cite{ishimoto2020jeffery}. 
In index notation, we find 
\begin{align}
    R^{FE}_{ijk}
    &=
    D^\text{n}
    (\hat{e}_{i}\hat{e}_{j}\hat{e}_{k}
    -\hat{e}_i\delta_{jk}/3
    )
    +
    D^\text{t}
    \left[
    (
    \delta_{ij}-\hat{e}_{i}\hat{e}_{j})
    e_k
    +
    (
    \delta_{ik}-\hat{e}_{i}\hat{e}_{k})
    e_j
    \right]
    +
    D^\text{o}
    (
    \epsilon_{ij\ell}\hat{e}_\ell\hat{e}_k
    +
    \epsilon_{ik\ell}\hat{e}_\ell\hat{e}_j
    ),
    \\
    R^{TE}_{ijk}
    &=
    E^\text{n}
    (\hat{e}_{i}\hat{e}_{j}\hat{e}_{k}
    -\hat{e}_i\delta_{jk}/3
    )
    +
    E^\text{t}
    \left[
    (
    \delta_{ij}-\hat{e}_{i}\hat{e}_{j})
    e_k
    +
    (
    \delta_{ik}-\hat{e}_{i}\hat{e}_{k})
    e_j
    \right]
    +
    E^\text{o}
    (
    \epsilon_{ij\ell}\hat{e}_\ell\hat{e}_k
    +
    \epsilon_{ik\ell}\hat{e}_\ell\hat{e}_j
    )
    ,
    \label{eq:FETE}
\end{align}
for the shear-induced resistance tensors, and
\begin{align}
    R^{SV}_{ijk}
    &=
    F^\text{n}
    (\hat{e}_{i}\hat{e}_{j}\hat{e}_{k}
    -\hat{e}_k\delta_{ji}/3
    )
    +
    F^\text{t}
    \left[
    (
    \delta_{kj}-\hat{e}_{k}\hat{e}_{j})
    e_i
    +
    (
    \delta_{ik}-\hat{e}_{i}\hat{e}_{k})
    e_j
    \right]
    +
    F^\text{o}
    (
    \epsilon_{kj\ell}\hat{e}_\ell\hat{e}_i
    +
    \epsilon_{ki\ell}\hat{e}_\ell\hat{e}_j
    ),
    \\
    R^{S\Omega}_{ijk}
    &=
    G^\text{n}
    (\hat{e}_{i}\hat{e}_{j}\hat{e}_{k}
    -\hat{e}_k\delta_{ji}/3
    )
    +
    G^\text{t}
    \left[
    (
    \delta_{kj}-\hat{e}_{k}\hat{e}_{j})
    e_i
    +
    (
    \delta_{ik}-\hat{e}_{i}\hat{e}_{k})
    e_j
    \right]
    +
    G^\text{o}
    (
    \epsilon_{kj\ell}\hat{e}_\ell\hat{e}_i
    +
    \epsilon_{ki\ell}\hat{e}_\ell\hat{e}_j
    )
    ,
    \label{eq:SVSO}
\end{align}
for the velocity-induced resistance tensors. 
In the above, the superscript ``n'' denotes the drag parallel to the axis, ``t'' denotes that transverse to it, and ``o'' denotes the anti-symmetric drag.

Under the cylindrical symmetry, a rank-4 tensor is allowed to have five coefficients~\cite{khain2022}.
In component notation, it is given by
\begin{align}
    R^{SE}_{ijk\ell}
    &=
    H^\text{tt}
    [
     (\delta_{ik}-\hat{e}_i \hat{e}_k)
      (\delta_{j\ell}-\hat{e}_j \hat{e}_\ell)
    +
    (\delta_{i\ell}-\hat{e}_i \hat{e}_\ell)
      (\delta_{jk}-\hat{e}_j \hat{e}_k)
    -
    (\delta_{ij}-\hat{e}_i \hat{e}_j)
      (\delta_{k\ell}-\hat{e}_k \hat{e}_\ell)
    ]
    \nonumber\\
    &
    +
    H^\text{nn}
    \left[
    3
    e_ie_je_ke_\ell
    -
    (
    e_ie_j\delta_{k\ell}
    +
    \delta_{ij}e_ke_\ell
    )
    +
    \delta_{ij}\delta_{k\ell}/3
    \right]
    \nonumber\\
    &+
    H^\text{nt}
    [
    e_ie_k
    (\delta_{j\ell}-e_je_\ell)
    +
    e_ie_\ell(\delta_{jk}-e_je_k)
    +
    (\delta_{ik}-e_ie_k)e_je_\ell
    +
    (\delta_{i\ell}-e_ie_\ell)e_je_k
    ]
    \nonumber\\
    &+
    H^\text{ott}
    \hat{e}_q
    [
    \epsilon_{qik}
    (\delta_{\ell j}-\hat{e}_\ell \hat{e}_j)
    +
    \epsilon_{qi\ell}
    (\delta_{kj}-\hat{e}_k\hat{e}_j)
    +
    \epsilon_{qjk}
    (\delta_{\ell i}-\hat{e}_\ell \hat{e}_i)
    +
    \epsilon_{qj\ell}
    (\delta_{ki}-\hat{e}_k\hat{e}_i)
    ]/2
    \nonumber\\
    &-
    H^\text{ont}
    \hat{e}_q
    [
    \epsilon_{qik}
    \hat{e}_\ell \hat{e}_j
    +
    \epsilon_{qi\ell}
    \hat{e}_k \hat{e}_j
    +
    \epsilon_{qjk}
    \hat{e}_\ell \hat{e}_i
    +
    \epsilon_{qj\ell}
    \hat{e}_k \hat{e}_i
    ]
    \label{eq:SECartesian}
    .
\end{align}

Further symmetry considerations can simplify the rank-3 coupling tensors.
Components that are invariant under reflection containing the $\mathbf{\hat{e}}$-axis are parity-even, while those that change their sign are parity-odd.
Noting that the force transforms as $(F_x,F_y,F_z)\to(-F_x,F_y,F_z)$ under the mirror reflection $(x,y,z)\to(-x,y,z)$, the coefficients $D^\text{n}$ and $D^\text{t}$ are invariant under this transformation, while $D^\text{o}$ is not.
Given that $D^\text{n}$ and $D^\text{t}$ are parity-even and $D^\text{o}$ is parity-odd and from the Onsager-Casimir reciprocity, one can conclude that $D^\text{n}=F^\text{n}, D^\text{t}=F^\text{t},$ and $-D^\text{o}=F^\text{o}$.
The transposed counterpart of the force-strain tensor therefore takes the form
\begin{align}
    R^{SV}_{ijk}
    &=
    D^\text{n}
    (\hat{e}_{i}\hat{e}_{j}\hat{e}_{k}
    -\hat{e}_k\delta_{ji}/3
    )
    +
    D^\text{t}
    \left[
    (
    \delta_{kj}-\hat{e}_{k}\hat{e}_{j})
    e_i
    +
    (
    \delta_{ik}-\hat{e}_{i}\hat{e}_{k})
    e_j
    \right]
    -D^\text{o}
    (
    \epsilon_{kj\ell}\hat{e}_\ell\hat{e}_i
    +
    \epsilon_{ki\ell}\hat{e}_\ell\hat{e}_j
    )
    .
\end{align}
Noting that the torque transforms as $(T_x,T_y,T_z)\to(T_x,-T_y,-T_z)$, one can find $E^\text{n}=-G^\text{n},E^\text{t}=-G^\text{t},$ and $E^\text{o}=G^\text{o}$.
Then the transposed counterpart of the torque-strain tensor reduces to
\begin{align}
    R^{S\Omega}_{ijk}
    &=
    -
    E^\text{n}
    (\hat{e}_{i}\hat{e}_{j}\hat{e}_{k}
    -\hat{e}_k\delta_{ji}/3
    )
    -
    E^\text{t}
    \left[
    (
    \delta_{kj}-\hat{e}_{k}\hat{e}_{j})
    e_i
    +
    (
    \delta_{ik}-\hat{e}_{i}\hat{e}_{k})
    e_j
    \right]
    +
    E^\text{o}
    (
    \epsilon_{kj\ell}\hat{e}_\ell\hat{e}_i
    +
    \epsilon_{ki\ell}\hat{e}_\ell\hat{e}_j
    )
    .
\end{align}

By using the five-component symmetric and traceless basis in Eq.~\eqref{eq:tau1-5}, the rank-3 and -4 tensors above can be presented in matrix notation.
The coupling components read
\begin{align}
    \tilde{\mathbf{R}}^{FE}
    &=
    \sqrt{2}
       \begin{pmatrix}
           0 & 0 & 0  & D^\text{o} & D^\text{t} \\
           0 & 0 & 0 & D^\text{t} &-D^\text{o} \\
           0 & 0 & D^\text{n}/\sqrt{3} & 0 & 0 
       \end{pmatrix}
       ,\quad
       \tilde{\mathbf{R}}^{TE}
    =
    \sqrt{2}
       \begin{pmatrix}
           0 & 0 & 0  & E^\text{o} & E^\text{t} \\
           0 & 0 & 0 & E^\text{t} &-E^\text{o} \\
           0 & 0 & E^\text{n}/\sqrt{3} & 0 & 0 
       \end{pmatrix}
       ,
\\
    \tilde{\mathbf{R}}^{SV}
    &=
    \sqrt{2}
       \begin{pmatrix}
           0 & 0 & 0   \\  
           0 & 0 & 0   \\
           0 & 0 & D^\text{n}/\sqrt{3}  \\
           -D^\text{o} & D^\text{t} &  0\\ 
           D^\text{t} & D^\text{o} & 0
       \end{pmatrix}
       ,\quad
       \tilde{\mathbf{R}}^{S\Omega}
    =
    \sqrt{2}
       \begin{pmatrix}
           0 & 0 & 0   \\
           0 & 0 & 0   \\
           0 & 0 & -E^\text{n}/\sqrt{3}  \\
           E^\text{o} & -E^\text{t} &  0\\ 
           -E^\text{t} & -E^\text{o} & 0
       \end{pmatrix}       
       ,
       \label{eq:RFESOtau}
\end{align}
and the stresslet-strain tensor becomes
\begin{align}
       \tilde{\mathbf{R}}^{SE}
       =
       2
       \begin{pmatrix}
           H^\text{tt} & H^\text{ott} & 0  & 0 & 0 \\
           -H^\text{ott} & H^\text{tt} & 0 & 0 &0 \\
           0 & 0 & H^\text{nn} & 0 & 0 \\
           0 & 0 & 0 & H^\text{nt} & H^\text{ont} \\
           0 & 0 & 0 & -H^\text{ont} & H^\text{nt}\\
       \end{pmatrix}
       \label{eq:SEtau}
    .
\end{align}

\pagebreak

\subsection{Resistance tensors of a cone with pseudo-scalar odd slip}

Here we provide the numerical values of the resistance tensor of a cone, computed using the Boundary Element Method (BEM) adapted from the BEMLIB library  \cite{pozrikidis2002}. 
A right circular cone with axis along the direction $\mathbf{\hat{p}}$ has a base radius $a=1$ and a height $h=2$. In the following, we also set $\eta=1$. 
The base of the cone is located at $(0,0,-0.476)$, chosen such that the center of reaction coincides with the origin.
By setting $\mathbf{\hat{p}}$ along the $z$-axis, without loss of generality, the translational and rotational part of the grand resistance tensor of a no-slip cone is numerically obtained as
\begin{align}
    \begin{pmatrix}
        \mathbf{R}^{FV}_\text{NS} & \mathbf{R}^{F\Omega}_\text{NS} \\
        \mathbf{R}^{TV}_\text{NS} & \mathbf{R}^{T\Omega}_\text{NS}
    \end{pmatrix}
    =
\begin{pmatrix}
16.38 & 0 & 0 & 0 & 0 & 0 \\
0 & 16.38 & 0 & 0 & 0 & 0 \\
0 & 0 & 17.00 & 0 & 0 & 0 \\
0 & 0 & 0 & 19.81 & 0 & 0 \\
0 & 0 & 0 & 0 & 19.81 & 0 \\
0 & 0 & 0 & 0 & 0 & 16.31
\end{pmatrix}
\label{eq:RFVcone0}
.
\end{align}
For the resistance tensors involving shear and stresslet moments, we use the 5-component basis introduced in the previous section. 
The resulting matrices are 
\begin{align}
    \mathbf{\tilde{R}}^{FE}_\text{NS}
    =
    \begin{pmatrix}
0 & 0 & 0 & 0 & 1.64 \\
0 & 0 & 0 & 1.64 & 0 \\
0 & 0 & -2.67 & 0 & 0
\end{pmatrix}
,\quad
    \mathbf{\tilde{R}}^{TE}_\text{NS}
    =
    \begin{pmatrix}
0 & 0 & 0 & 0.14 & 0 \\
0 & 0 & 0 & 0 & -0.14 \\
0 & 0 & 0 & 0 & 0
\end{pmatrix}
    ,
    \quad
    \mathbf{\tilde{R}}^{SE}_\text{NS}
    =
    \begin{pmatrix}
13.14 & 0 & 0 & 0 & 0 \\
0 & 13.14 & 0 & 0 & 0 \\
0 & 0 & 14.29 & 0 & 0 \\
0 & 0 & 0 & 13.99 & 0 \\
0 & 0 & 0 & 0 & 13.99
\end{pmatrix}
    .
    \label{eq:SEcone0}
\end{align}

We compute the resistance tensor of the same cone with a pseudo-scalar odd slip boundary to linear order in $\lambda^\text{SO}$ using the perturbative expansion from Eq.~\eqref{eq:LRT_oPS}. 
As required by the Onsager-Casimir reciprocity~\eqref{eq:symmetryR}, the linear correction part can be written as
\begin{align}
    \bm{\mathsf{R}}
    =
    \bm{\mathsf{R}}_\text{NS}
    +
    \lambda^\text{SO}
    \begin{pmatrix}
        \bm{\mathcal{T}} & \bm{\mathcal{R}} & \bm{\mathcal{V}}\\
        -\bm{\mathcal{R}}^\top & \bm{\mathcal{U}}& \bm{\mathcal{X}} \\
         -\bm{\mathcal{V}}^\top & 
        -\bm{\mathcal{X}}^\top & \bm{\mathcal{W}}
    \end{pmatrix}
    .
\end{align}
By using the tractions obtained using the BEM method on the no-slip cone and numerically evaluating the integrals in Eq.~\eqref{eq:LRT_oPS}, we obtain
\begin{align}
    \begin{pmatrix}
                \bm{\mathcal{T}} & \bm{\mathcal{R}}  \\
        -\bm{\mathcal{R}}^\top & \bm{\mathcal{U}} \\
    \end{pmatrix}
    =
\begin{pmatrix}
0 & 2.11  & 0 & -21.5  & 0 & 0 \\
-2.11  & 0 & 0 & 0 & -21.5 & 0 \\
0 & 0 & 0 & 0 & 0 & -24.5 \\
21.5  & 0 & 0 & 0 & 2.78 & 0 \\
0 & 21.5  & 0 & -2.78 & 0 & 0 \\
0 & 0 & 24.5 & 0 & 0 & 0
\end{pmatrix}
,
\label{eq:RFVcone}
\end{align}
for the translational-rotational part and
\begin{align}
    \bm{\tilde{\mathcal{V}}}
    =
\begin{pmatrix}
0 & 0 & 0 & 1.27 & 0 \\
0 & 0 & 0 & 0 & -1.27 \\
0 & 0 & 0 & 0 & 0
\end{pmatrix}
    ,\quad
    \bm{\tilde{\mathcal{X}}}
    =
   \begin{pmatrix}
0 & 0 & 0 & 0 & -0.708 \\
0 & 0 & 0 &  -0.708 & 0 \\
0 & 0 & -9.87 & 0 & 0
\end{pmatrix}
    ,
    \quad
    \bm{\tilde{\mathcal{W}}}
    =
    \begin{pmatrix}
0 & 11.3 & 0 & 0 & 0 \\
- 11.3 & 0 & 0 & 0 & 0 \\
0 & 0 & 0 & 0 & 0 \\
0 & 0 & 0 & 0 & 12.0 \\
0 & 0 & 0 & -12.0 & 0
\end{pmatrix}
\label{eq:RSEcone}
,
\end{align}
for the shear components transformed into the $\bm\tau$ basis.

Comparing the numerical results in Eqs.~\eqref{eq:RFVcone0}-\eqref{eq:RSEcone} with the symmetry-compatible tensorial structures shown in Eqs.~\eqref{eq:RFVRTV},~\eqref{eq:RFESOtau}, and \eqref{eq:SEtau}, we find for an odd-slip SO cone with its axisymmetric axis $\mathbf{\hat{p}}$
\begin{align}
    \mathbf{R}^{FV}
    = 
    A^\text{t} \mathbf{P}^\perp
    +
    A^\text{n} \mathbf{\hat{p}\hat{p}}
    +
    \lambda^\text{SO}A^\text{o}\bm{\epsilon}\cdot\mathbf{\hat{p}},
    \quad
    \mathbf{R}^{T\Omega}
    = 
    B^\text{t} \mathbf{P}^\perp
    +
    B^\text{n} \mathbf{\hat{p}\hat{p}}
    +
    \lambda^\text{SO}B^\text{o}\bm{\epsilon}\cdot\mathbf{\hat{p}},
    \quad
    \mathbf{R}^{F\Omega}
    =
    \lambda^\text{SO}(
    C^\text{t} \mathbf{P}^\perp
    +
    C^\text{n}\mathbf{\hat{p}\hat{p}})
    ,
    \label{eq:Rcone1}
\end{align}
for the translational and rotational part, and 
\begin{align}
    R^{FE}_{ijk}
    &=
    D^\text{n}
    \hat{p}_i
    (\hat{p}_j\hat{p}_k
    -\delta_{jk}/3
    )
    +
    D^\text{t}
    (
    P^\perp_{ij}
    \hat{p}_k
    +
    P^\perp_{ik}
    \hat{p}_j
    )
    +
    \lambda^\text{SO}
    D^\text{o}
    (
    \epsilon_{ij\ell}\hat{p}_\ell\hat{p}_k
    +
    \epsilon_{ik\ell}\hat{p}_\ell\hat{p}_j
    )
    \nonumber
    ,\\
    R^{TE}_{ijk}
    &=
    \lambda^\text{SO}
    E^\text{n}
    \hat{p}_i
    (\hat{p}_j\hat{p}_k
    -\delta_{jk}/3
    )
    +
    \lambda^\text{SO}
    E^\text{t}
    (
    P^\perp_{ij}
    \hat{p}_k
    +
    P^\perp_{ik}
    \hat{p}_j
    )
    +
    E^\text{o}
    (
    \epsilon_{ij\ell}\hat{p}_\ell\hat{p}_k
    +
    \epsilon_{ik\ell}\hat{p}_\ell\hat{p}_j
    )
    ,
    \nonumber
    \\
    R^{SE}_{ijk\ell}
    &=
    H^\text{tt}
    (
    P^\perp_{ik}P^\perp_{j\ell}
    +
    P^\perp_{i\ell}P^\perp_{jk}
    -
    P^\perp_{ij}P^\perp_{k\ell}
    )
    \nonumber\\
    &+
    H^\text{nn}
    \left[
    3
    \hat{p}_i\hat{p}_j\hat{p}_k\hat{p}_\ell
    -
    (
    \hat{p}_i\hat{p}_j\delta_{k\ell}
    +
    \delta_{ij}\hat{p}_k\hat{p}_\ell
    )
    +
    \delta_{ij}\delta_{k\ell}/3
    \right]
    \label{eq:Rcone2}
     \\
    &+
    H^\text{nt}
    (
    \hat{p}_i\hat{p}_k
    P^\perp_{j\ell}
    +
    \hat{p}_i\hat{p}_\ell P^\perp_{jk}
    +
    P^\perp_{ik}\hat{p}_j\hat{p}_\ell
    +
    P^\perp_{i\ell}\hat{p}_j\hat{p}_k
    )
    \nonumber\\
    &+
    \lambda^\text{SO}
    H^\text{ott}
    \hat{p}_q
    [
    \epsilon_{qik}
    P^\perp_{\ell j}
    +
    \epsilon_{qi\ell}
    P^\perp_{kj}
    +
    \epsilon_{qjk}
    P^\perp_{\ell i}
    +
    \epsilon_{qj\ell}
    P^\perp_{ki}
    ]/2
    \nonumber\\
    &-
    \lambda^\text{SO}
    H^\text{ont}
    \hat{p}_q
    [
    \epsilon_{qik}
    \hat{p}_\ell \hat{p}_j
    +
    \epsilon_{qi\ell}
    \hat{p}_k \hat{p}_j
    +
    \epsilon_{qjk}
    \hat{p}_\ell \hat{p}_i
    +
    \epsilon_{qj\ell}
    \hat{p}_k \hat{p}_i
    ]
    ,
    \nonumber
\end{align}
for the other sub-tensors, where $\mathbf{P}^\perp=\mathbf{I}-\mathbf{\hat{p}\hat{p}}$ is a projection operator to the plane normal to the axis.
In the above, the drag coefficients of a pseudo-scalar cone have the values
\begin{align}
    A^\text{t} &= 16.38  ,&
    A^\text{n} &= 17.00 ,&
    A^\text{o} &= 2.11  ,&
    B^\text{t} &= 19.81 ,&
    B^\text{n} &= 16.31 ,&
    B^\text{o} &= 2.78 ,&
    C^\text{t} &= -21.5,&
    C^\text{n} &= -24.5,&\nonumber\\
    D^\text{t} &= 1.16 ,&
    D^\text{n} &= -3.27,&
    D^\text{o} &= 0.90 ,&
    E^\text{t} &= -0.50 ,&
    E^\text{n} &= -12.09,&
    E^\text{o} &= 0.10, &\nonumber\\
    H^\text{tt} &= 6.57 ,&
    H^\text{nn} &= 7.15,&
    H^\text{nt} &= 7.00,&
    H^\text{ott} &= 5.65,&
    H^\text{ont} &= 6.0.&
\end{align}

The odd-slip corrections to the no-slip part are specifically given by
\begin{align}
    \bm{\mathcal{T}}
    &=
    A^\text{o}
    \bm\epsilon\cdot\mathbf{\hat{p}}
    ,
    \nonumber\\
    \bm{\mathcal{U}}
    &=
    B^\text{o}
    \bm\epsilon\cdot\mathbf{\hat{p}}
    ,\nonumber\\
    \bm{\mathcal{R}}
    &=
    C^\text{t}
    \mathbf{P}^\perp
    +
    C^\text{n}
    \mathbf{\hat{p}\hat{p}},
    \nonumber\\
    \mathcal{V}_{ijk}
    &=
    D^\text{o}
    (
    \epsilon_{ij\ell}\hat{p}_\ell\hat{p}_k
    +
    \epsilon_{ik\ell}\hat{p}_\ell\hat{p}_j
    )
    ,
    \label{eq:Rcone}\\
    \mathcal{X}_{ijk}
    &=
    E^\text{n}
    \hat{p}_{i}
    \left(
    \hat{p}_{j}\hat{p}_{k}
    -
    \delta_{jk}/3
    \right)
    +
    E^\text{t}
    (
    P^\perp_{ij}
    \hat{p}_k
    +
    P^\perp_{ik}
    \hat{p}_j
    ),
    \nonumber
    \\
    \mathcal{W}_{ijk\ell}
    &=
    H^\text{ott}
    \hat{p}_q
    (
    \epsilon_{qik}
    P^\perp_{\ell j}
    +
    \epsilon_{qi\ell}
    P^\perp_{kj}
    +
    \epsilon_{qjk}
    P^\perp_{\ell i}
    +
    \epsilon_{qj\ell}
    P^\perp_{ki}
    )/2\nonumber\\
    &-
    H^\text{ont}
    \hat{p}_q
    (
    \epsilon_{qik}
    \hat{p}_\ell \hat{p}_j
    +
    \epsilon_{qi\ell}
    \hat{p}_k \hat{p}_j
    +
    \epsilon_{qjk}
    \hat{p}_\ell \hat{p}_i
    +
    \epsilon_{qj\ell}
    \hat{p}_k \hat{p}_i
    ).\nonumber
\end{align}
By setting $A^\text{o}=\mathcal{T}$, $ B^\text{o}=\mathcal{U}$, $C^\text{t}=\mathcal{R}_\perp$, $C^\text{n}=\mathcal{R}_\|$, $ D^\text{o}=\mathcal{V}$, $ E^\text{n}=\mathcal{X}_\|$, $ E^\text{t}=\mathcal{X}_\perp$, $ H^\text{ott}/2=\mathcal{W}_\perp$, and $H^\text{ont}=\mathcal{W}_\|$, the above expressions are reduced to Eq.~\eqref{eq:RSOconesub} in the main text.

\subsection{Particle stress in a suspension of cones}

Here we construct the constitutive relation of a dilute suspension of cones with pseudo-scalar odd slip boundaries.
The particles are force-free with a fixed orientation $\mathbf{\hat{p}}$, which requires a restoring torque perpendicular to that direction.
The particle stress can formally be written as
\begin{align}
    \boldsymbol{\Sigma}^\text{p}
    =
    n\left( \mathbf{S}
    -
    \frac{1}{2}
    \boldsymbol{\epsilon}\cdot
    \mathbf{T}\right).
    \label{eq:pstress}
\end{align}
The stresslet and torque acting on a particle are obtained by solving the matrix equation $[\bm{0},\mathbf{T},\mathbf{S}]= -\bm{\mathsf{R}} \cdot [\mathbf{V},\boldsymbol\Omega-\boldsymbol\Omega^\infty,-\mathbf{E}^\infty]^\top$ with the conditions, $\mathbf{T}\cdot\mathbf{\hat{p}}=0$ and $\bm{\Omega}\times\mathbf{\hat{p}}=\mathbf{0}$.
We now separate the velocity of a particle $\bm{v}=[V_x,V_y,V_z,\Omega_z]$ and the imposed flows $\bm{u}=[-\Omega^\infty_x, -\Omega^\infty_y, -\tilde{u}^{1},\dots,-\tilde{u}^{5}]$, with $\tilde{u}^\alpha=\tau^\alpha_{ij}\partial_jv_i^\infty$ being the transformed velocity gradient.
In this new basis, the matrix equation can be written as
\begin{align}
    \begin{pmatrix}
        \bm{0}\\
        \bm{s}
    \end{pmatrix}
    =
    -
    \begin{pmatrix}
        \bm{\zeta}_{fv} & \bm{\zeta}_{fu}\\
        \bm{\zeta}_{fu}^\ast &
         \bm{\zeta}_{su}
    \end{pmatrix}
        \begin{pmatrix}
        \bm{v}\\
        \bm{u}
    \end{pmatrix}
    ,
    \label{eq:ABBC}
\end{align}
where we have defined the $7$-component force-moment vector $\bm{s}=[T_x,T_y,\tilde{S}^1,\dots,\tilde{S}^5]$ with $\tilde{S}^\alpha=\tau^\alpha_{ij}S_{ij}$, and the asterisk symbol denotes the transposition of a matrix under the sign flip $\lambda^\text{SO}\to-\lambda^\text{SO}$.
In the above, the resistance matrices are given by
\begin{align}
    \bm{\zeta}_{fv}
    &=
\begin{pmatrix}
16.38 & 2.11\lambda^\text{SO} & 0 & 0 \\
-2.11\lambda^\text{SO} & 16.38 & 0 & 0 \\
0 & 0 & 17.00 & -24.5\lambda^\text{SO} \\
0 & 0 & 24.5\lambda^\text{SO} & 16.31
\end{pmatrix},
\\
    \bm{\zeta}_{fu}
    &=
\begin{pmatrix}
-21.5\lambda^\text{SO} & 0 & 0 & 0 & 0 & 1.27\lambda^\text{SO} & 1.64 \\
0 & -21.5\lambda^\text{SO} & 0 & 0 & 0 & 1.64 & -1.27\lambda^\text{SO} \\
0 & 0 & 0 & 0 & -2.67 & 0 & 0 \\
0 & 0 & 0 & 0 & -9.87\lambda^\text{SO} & 0 & 0
\end{pmatrix}
,\\
\bm{\zeta}_{su}
&=
\begin{pmatrix}
19.81 & 2.78\lambda^\text{SO} & 0 & 0 & 0 & 0.14 & -0.708\lambda^\text{SO} \\
-2.78\lambda^\text{SO} & 19.81 & 0 & 0 & 0 & -0.708\lambda^\text{SO} & -0.14 \\
0 & 0 & 13.14 & 11.3\lambda^\text{SO} & 0 & 0 & 0 \\
0 & 0 & -11.3\lambda^\text{SO} & 13.14 & 0 & 0 & 0 \\
0 & 0 & 0 & 0 & 14.29 & 0 & 0 \\
0.14 & 0.708\lambda^\text{SO} & 0 & 0 & 0 & 13.99 & 12.0\lambda^\text{SO} \\
0.708\lambda^\text{SO} & -0.14 & 0 & 0 & 0 & -12.0\lambda^\text{SO} & 13.99
\end{pmatrix}
.
\end{align}
The force moments can be formally obtained by expressing the solution of Eq.~\eqref{eq:ABBC} as $\bm{s}=\bm{\mathsf{Z}}\cdot\bm{u}$ with the drag matrix given by $\bm{\mathsf{Z}}=\bm{\zeta}_{fu}^\ast\cdot\bm{\zeta}_{fv}^{-1}\cdot\bm{\zeta}_{fu}-\bm{\zeta}_{su}$.
Expanding the resistance matrix up to linear order in the slip length yields
\begin{align}
    \bm{\mathsf{Z}}
    =
    \begin{pmatrix}
-19.81 & 0 & 0 & 0 & 0 & -0.14 & 0 \\
0 & -19.81 & 0 & 0 & 0 & 0 & 0.14 \\
0 & 0 & -13.14 & 0 & 0 & 0 & 0 \\
0 & 0 & 0 & -13.14 & 0 & 0 & 0 \\
0 & 0 & 0 & 0 & -13.87 & 0 & 0 \\
-0.14 & 0 & 0 & 0 & 0 & -13.83 & 0 \\
0 & 0.14 & 0 & 0 & 0 & 0 & -13.83
\end{pmatrix}
+\lambda^\text{SO}
\begin{pmatrix}
0 & -2.78 & 0 & 0 & 0 & 0 & 2.86 \\
2.78 & 0 & 0 & 0 & 0 & 2.86 & 0 \\
0 & 0 & 0 & -11.30 & 0 & 0 & 0 \\
0 & 0 & 11.30 & 0 & 0 & 0 & 0 \\
0 & 0 & 0 & 0 & 0 & 0 & 0 \\
0 & -2.86 & 0 & 0 & 0 & 0 & -12.23 \\
-2.86 & 0 & 0 & 0 & 0 & 12.23 & 0
\end{pmatrix}
.
\end{align}

From the definition of the particle stress in Eq.~\eqref{eq:pstress}, the effective viscosity of the suspension is expressed as an $8\times8$ matrix in the eight-component $\bm\tau$ basis.
Noting that $\bm{\Omega}^\infty=\frac{1}{2}\nabla\times\mathbf{v}^\infty$,
\begin{align}
    \bm{\tilde{\eta}}^\text{eff}
    =
    2\eta
    \begin{pmatrix}
0 & 0 & 0 & 0 & 0 & 0 & 0 & 0 \\
0 & 0 & 0 & 0 & 0 & 0 & 0 & 0 \\
0 & 0 & 0 & 0 & 0 & 0 & 0 & 0 \\
0 & 0 & 0 & 1 & 0 & 0 & 0 & 0 \\
0 & 0 & 0 & 0 & 1 & 0 & 0 & 0 \\
0 & 0 & 0 & 0 & 0 & 1 & 0 & 0 \\
0 & 0 & 0 & 0 & 0 & 0 & 1 & 0 \\
0 & 0 & 0 & 0 & 0 & 0 & 0 & 1
\end{pmatrix}
+
\eta\phi&
   \begin{pmatrix}
4.73 & 0 &0 & 0 & 0 & 0 & -0.05 & 0 \\
0 & 4.73 & 0& 0 & 0 & 0 & 0 & 0.05 \\
0 & 0 & 0 & 0 &0& 0 & 0 & 0 \\
0 & 0 & 0 &6.27 & 0 & 0 & 0 & 0 \\
0 & 0 & 0 & 0 & 6.27 & 0 & 0 & 0 \\
0 & 0 & 0 & 0 & 0 & 6.62 & 0 & 0 \\
-0.05 & 0 & 0& 0 & 0 & 0 & 6.60 & 0 \\
0 & 0.05 & 0 & 0&  0 & 0 & 0 & 6.60
\end{pmatrix}
\nonumber\\
+
\eta\phi(\lambda^\text{SO}/a)
&
\begin{pmatrix}
0 & 0.66 & 0 & 0 & 0 & 0 & 0 & 0.97\\
-0.66 & 0 & 0 &0 & 0 & 0 & 0.97 & 0 \\
0 & 0 & 0 & 0 &0& 0 & 0 & 0 \\
0 & 0 & 0 & 0& 5.40 & 0 & 0 & 0 \\
0 & 0 & 0& -5.40 & 0 & 0 & 0 & 0 \\
0 & 0 & 0 &0 & 0 & 0 & 0 & 0 \\
0 & -0.97 & 0& 0 & 0 & 0 & 0 & 5.84 \\
-0.97 & 0 & 0& 0 & 0 & 0 & -5.84 & 0
\end{pmatrix}
.
\end{align}
where $\phi=\frac{1}{3}\pi ha^2n$ denotes the volume fraction of the cones. In the final expression, we have re-introduced the physical variables $\eta$ and $a$ while keeping the ratio $h/a=2$. 
In contrast to the suspension of VO-slip spheres, the vectorial shape of suspended cones gives rise to two odd viscosities.
Further, the shear-rotational coupling acquires an odd response.

\section{Shear-induced stresslet of a Janus sphere with pseudo-scalar odd slip}
\label{app:Janus}
We examine here a Janus sphere with one half (hemispherical surface $\mathcal{S}_1$) exhibiting the pseudo-scalar odd slip and the other half with a no-slip boundary.
The polar direction is denoted by a unit vector $\mathbf{\hat{p}}$.
From the Lorentz reciprocal theorem, Eq.~\eqref{eq:VRV}, the resistance matrix follows
\begin{align}
    \bm{\mathbf{V}}_\text{NS}
    \cdot
    (
    \bm{\mathbf{R}}
    -
    \bm{\mathbf{R}}_\text{NS})
    \cdot\bm{\mathbf{V}}
    =
    -
    \frac{\lambda^\text{SO}}{\eta}
    \int_{\mathcal{S}_1}dS\,
    (\mathbf{\hat{n}}\times\mathbf{f})\cdot\mathbf{f}_\text{NS}
    .
\end{align}
Here the main problem consists of the Janus sphere in an imposed flow $\mathbf{E}^\infty \cdot\mathbf{r}$ and the auxiliary problem of a NS sphere of the same instantaneous shape in a background flow $\mathbf{E}^\infty_\text{NS}\cdot\mathbf{r}$.
To first order in $\lambda^\text{SO}$, the corrections read
\begin{align}
\mathbf{E}^\infty_\text{NS}:\mathbf{R}^{SE} 
    :\mathbf{E}^\infty 
    =
    \mathbf{E}^\infty:\mathbf{R}^{SE}_\text{NS} 
    :\mathbf{E}^\infty_\text{NS}
    -
    \eta
    \lambda^\text{SO}
    \int_{\mathcal{S}_1}dS\,
    [\mathbf{\hat{n}}\times
    (\mathbf{d}^E:\mathbf{E}^\infty )
    ]
    \cdot
    (
    \mathbf{d}^E
    :
    \mathbf{E}^\infty_\text{NS}
    )
    .
\end{align}
Eliminating the arbitrary tensors, the resistance matrix can be solved separately to obtain
\begin{align}
    R_{ijk\ell}^{SE}
    =
    R_{\text{NS},ijk\ell}^{SE}
    -
    \eta
    \lambda^\text{SO}
    \int_{\mathcal{S}_1}dS\,
    \epsilon_{mnp}\hat{n}_n
    d^E_{pk\ell}
    d^E_{mij}
    .
\end{align}
The integral can be evaluated to give the stresslet-strain tensor, which is composed of symmetric and anti-symmetric parts under the permutation of indices $ij\leftrightarrow k\ell$:
\begin{align}
    R_{ijk\ell}^{SE}
    =
    R_{\text{NS},ijk\ell}^{SE}
    +
    \frac{25}{16}
    \pi\eta a^3
    (\lambda^\text{SO}/a)
     &\hat{p}_q
    (
    \epsilon_{qik}P^\perp_{\ell j}
    +
    \epsilon_{qi\ell}P^\perp_{kj}
    +
    \epsilon_{qjk}P^\perp_{\ell i}
    +
    \epsilon_{qj\ell}P^\perp_{ki}
    )    ,
\end{align}
with $\mathbf{P}^\perp =\mathbf{I-\hat{p}\hat{p}}$ denoting a projection operator to the plane normal to the axis.
The first symmetric contribution leads to the classical effective shear viscosity (Einstein viscosity) in a suspension. 
In contrast, the second anti-symmetric term gives rise to odd viscosity.

\end{widetext}

\bibliography{myref}

\end{document}